\documentclass[12pt]{iopart}
\usepackage{bm}
\usepackage{iopams}

\usepackage{graphicx,epsfig}

\newcommand{\text}[1]{\mbox{\scriptsize{#1}}}

\begin{document}

\title{Phase transitions in a mechanical system coupled to Glauber spins}
\author{A Prados$^1$, L L Bonilla$^2$ and A Carpio$^3$}
\address{$^1$F\'{\i}sica Te\'{o}rica, Universidad de Sevilla,
Apartado de Correos 1065, E-41080, Sevilla, Spain}
\address{$^2$ G. Mill\'an Institute for Fluid Dynamics, Nanoscience and Industrial Mathematics, Universidad Carlos III de Madrid, 28911 Legan\'es, Spain}
\address{$^3$ Departamento de Matem\'atica Aplicada, Universidad Complutense de Madrid, 28040 Madrid, Spain}
\eads{\mailto{prados@us.es},\mailto{bonilla@ing.uc3m.es},\mailto{carpio@mat.ucm.es}}

\begin{abstract}
A harmonic oscillator linearly coupled with a linear chain of Ising spins is investigated. The $N$ spins in the chain interact with their nearest neighbours with a coupling constant proportional to the oscillator position and to $N^{-1/2}$, are in contact with a thermal bath at temperature $T$, and evolve under Glauber dynamics. The oscillator position is a stochastic process due to the oscillator-spin interaction which produces drastic changes in the equilibrium behaviour and the dynamics of the oscillator. Firstly, there is a second order phase transition at a critical temperature $T_c$ whose order parameter is the oscillator stable rest position: this position is zero above $T_c$ and different from zero below $T_c$. This transition appears because the oscillator moves in an effective potential equal to the harmonic term plus the free energy of the spin system at fixed oscillator position. Secondly, assuming fast spin relaxation (compared to the oscillator natural period), the oscillator dynamical behaviour is described by an effective equation containing a nonlinear friction term that drives the oscillator towards the stable equilibrium state of the effective potential. The analytical results are compared with numerical simulation throughout the paper.
\end{abstract}


\maketitle

\section{Introduction}
\label{s1}
Many physical processes are interpreted in terms of an oscillator coupled
to a thermal bath or to spin systems. Examples abound, mass spectrometry through a nanoelectromechanical oscillator whose resonant frequency decreases as single molecules are added to it \cite{boi09}, a spin representing a two-level system is coupled to a boson bath (the spin-boson system) to analyze loss of quantum coherence due to the bath \cite{leg87}, a classical oscillator coupled to a spin causes wave function collapse thereof \cite{bon92}, the classical version of the spin-phonon system describes the collective Jahn-Teller effect \cite{fed73,rik78}, large spin systems (single molecule magnets or nuclear spins) are coupled to a boson bath \cite{hic08}, etc.

In this work, we consider a mechanical degree of freedom represented by a classical harmonic oscillator coupled to a linear chain of $N$ Ising spins $\sigma_i$ ($i=1,\ldots, N$, $\sigma_i=\pm 1$) in contact with a thermal bath at temperature $T$. The energy of the combined system is equal to the energy of the oscillator alone plus a coupling term proportional to the oscillator position and to $N^{-1/2}\sum_{i=1}^N\sigma_i\sigma_{i+1}$. The spins flip stochastically according to Glauber dynamics \cite{Gl63}. As a consequence of the coupling, the oscillator equations of motion become stochastic, and both the position and the momentum of the oscillator become stochastic processes. The aim of this work is to understand how the equilibrium and the dynamics of the oscillator is affected by the interaction with the spin system (and vice versa).

The plan of the paper is as follows. The oscillator-spin model is described in section \ref{s2}. Its time evolution is governed by Newton's second law for the oscillator and the above mentioned Glauber dynamics for the spins. They can be put together in an evolution equation for the joint probability density of finding at time $t$ the oscillator at given values of its position and momentum and the Ising system at a given configuration. The canonical distribution at temperature $T$ is the equilibrium joint probability density. By summing over all possible spin configurations, we obtain the equilibrium probability density for the oscillator. The latter is a canonical distribution with an effective potential energy which is the sum of the harmonic potential and the free energy of the spin chain for fixed oscillator position.

Section \ref{s3} is devoted to analyzing the equilibrium configuration. By finding the minima of the effective potential, we show that there is a second order phase transition at a critical temperature $T_c$, with the stable rest position of the oscillator (equilibrium) as its order parameter. For $T>T_c$, the oscillator equilibrium position is the same as that of the uncoupled oscillator. For $T<T_c$, two symmetric nonzero equilibrium positions issue forth from zero as in the diagram of a pitchfork bifurcation. These nonzero equilibrium positions behave as $\sqrt{N}$ as the number of spins $N$ goes to infinity, so that the harmonic contribution to the energy be extensive in the thermodynamic limit. On the other hand, the fluctuations scale as $N^{-1/2}$ far from the critical temperature. Very close to $T_c$, there is a crossover and equilibrium fluctuations scale as $N^{-1/4}$. Although Ising spins in the chain are coupled to their nearest neighbours, their coupling constant is proportional to the oscillator position which makes their interaction effectively long range. Similar hidden long-range effective correlations that enable possible 1d phase transitions are present in biophysical systems. An example is DNA melting \cite{Pe06} which has been modeled by means of modified Ising systems \cite{DyP93,WyB85}, different from the one considered here.

The dynamics of the system is studied in section \ref{s4}. In the limit of fast relaxation of the spins compared to the natural period of the oscillator, there is a clear separation of time scales, a fast one associated to the relaxation of the spins and a slow one associated to the oscillator. In this regime, we find a reduced dynamics of the oscillator with nonlinear friction and a nonlinear force term. This nonlinear evolution equation is one of the main results of our paper. Basically, the spins approach their equilibrium distribution corresponding to the instantaneous value of the oscillator position. This produces the effective potential (already found in the equilibrium analysis) for the oscillator and gives rise to the nonlinear force term in its nonlinear evolution equation. On the other hand, the nonlinear friction is a purely dynamical effect that cannot be obtained from analyzing the equilibrium distribution of the system. This friction arises from the slow evolution of the oscillator resulting in a slight deviation of the Ising spins from its equilibrium with a fixed position of the oscillator. The friction term drives the system to equilibrium in the long time limit.  The stationary solutions of the reduced dynamics coincide with the oscillator equilibrium positions at any given temperature. We also discuss the expected range of validity of the nonlinear dynamical equation. For $T>T_c$ (section \ref{s4a}), we can linearize the oscillator reduced evolution equation about its stable rest state. The solutions are underdamped oscillations whose frequency decreases as $T$ decreases: the oscillator is slowed down by the spins. There is a narrow region of overdamped oscillations for temperatures very close to $T_c$.  A similar analysis is carried out in section \ref{s4b}, but for $T<T_c$. There is also a very narrow region of overdamped oscillations near $T_c$. For lower temperatures $T<T_c$, our theory predicts underdamped oscillations around one of the two nonvanishing stable equilibrium points.

In Section \ref{s5}, we compare numerical simulations for the model with the theoretical results and test the range of validity of the theory. The numerical simulations show excellent agreement with the theory for sufficiently high temperature $T>T_c$ and for 1000 or more spins. As $T$ decreases towards $T_c$, the numerical solutions of our theory and the simulations show the same qualitative trends, i.e., underdamped oscillations, but these oscillations are shifted by some constant value. This is an effect due to the initial conditions as reducing the size thereof brings again quantitative agreement between theory and simulations. Below the critical temperature, but not very far from it, more spins are necessary to attain good average values and our theory still gives an adequate description of the dynamical evolution of the system. As the temperature is further lowered, we again need fewer spins to attain good averages over spin indices and trajectories but there are qualitative differences between theory and simulations. Breakdown of the theoretical predictions is expected for sufficiently low temperatures, because of the divergence of the relaxation time of the spins \cite{Re80,ByP93b,ByP96}. Lastly, section \ref{s6} contains final remarks and comments.

\section{The model}
\label{s2}

We consider a system comprising one dimensional harmonic oscillator (mass $m$, frequency $\omega_0$, position $x$ and momentum $p$) and $N\gg 1$ internal degrees of freedom modeled by Ising spins ($\sigma_i=\pm1$, $i=1,\ldots,N$) in contact with a heat bath at temperature $T$.
The system has an energy
\numparts
\begin{eqnarray}
{\cal H}(x,p,\bm{\sigma}) & = & {\cal H}_0(x,p)+{\cal H}_{\text{int}}(x,\bm{\sigma}) \, \label{2.2a}\\
{\cal H}_0(x,p) &= &\frac{p^2}{2m}+\frac{1}{2}m\omega_0^2 x^2 \, , \label{2.2b}\\
{\cal H}_{\text{int}}(x,\bm{\sigma})&=&-\mu x \sum_{i=1}^N \sigma_i \sigma_{i+1}   \, , \label{2.2c}
\end{eqnarray}
\endnumparts
in which ${\cal H}_0(x,p)$ and ${\cal H}_{\text{int}}(x,\bm{\sigma})$ are the energy of the uncoupled oscillator and the interaction energy between the oscillator and the spins, respectively. The latter can also be understood as a nearest neighbour interaction between the spins with a coupling constant $J_{\text{eff}}$ which is proportional to the oscillator position $x$,
\begin{equation}\label{2.1}
  J_{\text{eff}}=\mu x \, .
\end{equation}
The parameter $\mu$ measures the strength of the coupling between the oscillator and the Ising system. Because of the sum over spins in (\ref{2.2c}), $\mu$ should decrease with $N$ for the system to have a well defined behaviour in the limit $N\rightarrow\infty$. We will show later that $\mu=\mu_0/\sqrt{N}$ as mentioned in section \ref{s1}. Alternatively, the  Hamiltonian (\ref{2.2a}) can also be written as
\numparts
\begin{eqnarray}
{\cal H}(x,p,\bm{\sigma}) & = & \frac{p^2}{2m}+{\cal V}(x,\bm{\sigma}) \, , \label{2.2d}\\
  {\cal V}(x,\bm{\sigma})&=&\frac{1}{2}m\omega_0^2 x^2- \mu x \sum_{i=1}^N \sigma_i \sigma_{i+1} \, , \label{2.2e}
\end{eqnarray}
where ${\cal V}(x,\bm{\sigma})$ is the total potential acting on the oscillator.

The dynamics of the system is governed by Hamilton's equations of motion
for the oscillator,
\numparts
\begin{eqnarray}
\dot{x} &=&\frac{p}{m}  \label{2.3a} \\
\dot{p}& =&-\frac{\partial\cal V}{\partial x} = -m \omega_0^2 x+\mu \sum_{i=1}^N \sigma_i \sigma_{i+1} \, , \label{2.3b}
\end{eqnarray}
or, equivalently,
\begin{equation}\label{2.3c}
    \ddot{x}+\omega_0^2 x=\frac{\mu}{m} \sum_{i=1}^N \sigma_i \sigma_{i+1} \, ,
\end{equation}
\endnumparts
and by an appropriate stochastic dynamics for the spins (because they are in contact with a heat bath at temperature $T$). For the sake of simplicity, the spins will be assumed to evolve with Glauber-like one spin flip dynamics. At any time $t$, the system may experience a transition from $(x,p,\bm{\sigma})$ to $(x,p,R_i\bm{\sigma})$ with a rate given by \cite{Gl63}
\begin{equation}\label{2.4}
W_i(\bm{\sigma}|x,p)=\frac{\alpha}{2} \left[
1-\frac{\gamma}{2} \sigma_{i} (\sigma_{i-1}+\sigma_{i+1}) \right],
\end{equation}
where $R_i\bm{\sigma}$ is the configuration obtained from $\bm{\sigma}$ by rotating the $i$-th spin. Here
\begin{equation}\label{2.5}
\gamma=\tanh\left( \frac{2 J_{\text{eff}}}{k_{B}T}\right)
=\tanh\left( \frac{2 \mu x}{k_{B}T}\right),
\end{equation}
$k_{B}$ is the Boltzmann constant and $T$ is the temperature of the
system. The
quantity $\alpha$ determines the characteristic attempt rate for the transitions in the Ising system.

In this way, the joint probability ${\cal P}(x,p, \bm{\sigma},t)$ of finding the oscillator with position $x$ and momentum $p$, and the spins in a configuration $\bm{\sigma}=\{\sigma_1,\sigma_2,\ldots,\sigma_N\}$ at time t obeys the  Liouville-master equation
\begin{eqnarray}\label{2.6}
\fl \partial_t {\cal P}(x,p,\bm{\sigma},t)+\frac{p}{m} \partial_x {\cal P}(x,p,\bm{\sigma},t)+ \left(-m \omega_0^2 x+\mu \sum_{i=1}^n \sigma_i \sigma_{i+1}\right) \partial_p {\cal P}(x,p,\bm{\sigma},t) \nonumber \\=\sum_{i=1}^N \left[ W_i(R_i\bm{\sigma}|x,p) {\cal P}(x,p,R_i\bm{\sigma},t)-
W_i(\bm{\sigma}|x,p){\cal P}(x,p,\bm{\sigma},t) \right] \, .
\end{eqnarray}
The equilibrium solution of this equation is the canonical distribution
\begin{equation}\label{2.7}
    {\cal P}_{\text{eq}}(x,p,\bm{\sigma})= \frac{1}{Z} e^{-\beta {\cal H}(x,p,\bm{\sigma})} \, ,
\end{equation}
where $Z$ is the partition function
\begin{equation}\label{2.8}
    Z=\int_{-\infty}^{+\infty} \rmd x \int_{-\infty}^{+\infty} \rmd p \sum_{\bm{\sigma}} e^{-\beta {\cal H}(x,p,\bm{\sigma})} \, ,
\end{equation}
and $\beta=(k_B T)^{-1}$. Since we are mainly interested in the behaviour of the oscillator, it will be useful to consider the marginal probability ${\cal P}_{\text{eq}}(x,p)$
\begin{equation}\label{2.9}
    {\cal P}_{\text{eq}}(x,p)=\sum_{\bm{\sigma}} {\cal P}_{\text{eq}}(x,p,\bm{\sigma})=\frac{1}{Z} e^{-\beta {\cal H}_0(x,p)} Z_{\text{Ising}}(x) \,
\end{equation}
where
\begin{equation}\label{2.10}
 Z_{\text{Ising}}(x)\equiv \rme^{-\beta {\cal F_{\text{Ising}}}} =\sum_{\bm{\sigma}} e^{-\beta {\cal H}_{\text{int}}(x,\bm{\sigma})}=\left[ 2 \cosh\left(\frac{J_{\text{eff}}}{k_B T}\right) \right]^N \,
\end{equation}
is the partition function of a 1d nearest neighbour Ising model with coupling constant $J_{\text{eff}}$, which depends on $x$ as given by (\ref{2.1}), and ${\cal F}_{\text{Ising}}(x)$ the corresponding free energy.  Therefore, ${\cal P}_{\text{eq}}(x,p)$ is readily rewritten,
\begin{equation}\label{2.11}
    {\cal P}_{\text{eq}}(x,p)=\frac{1}{Z} \exp \left\{-\beta \left[ \frac{p^2}{2m}+{\cal V}_{\text{eff}}(x) \right] \right\} \, ,
\end{equation}
with
\begin{eqnarray}\label{2.12}
    {\cal V}_{\text{eff}}(x) &=&\frac{1}{2}m\omega_0^2 x^2+{\cal F}_{\text{Ising}}(x)\, ,
    \\
\label{2.13}
    &=&\frac{1}{2}m \omega_0^2 x^2-N k_B T \left[\ln\cosh \left(\frac{\mu x}{k_B T}\right)+\ln 2\right] \, .
\end{eqnarray}
Equation (\ref{2.11}) suggests that ${\cal V}_{\text{eff}}$ is the effective potential acting on the oscillator due to its coupling to the $N$ Glauber spins. This point will be confirmed when the dynamics be analyzed in section \ref{s4}.

\subsection{Orders of magnitude and nondimensional equations}

\begin{table}[ht]
\begin{center}\begin{tabular}{ccccccc}
 \hline
 $x$ & $p$ &${\cal P}$ & $W_i$ & $t$ & $\epsilon$ &$\theta$  \\
$\frac{\mu N}{m\omega_0^2}$ & $\frac{\mu N}{\omega_0}$ & $\frac{m\omega_0^3}{\mu^2N^2}$ &$\alpha$ & $\frac{1}{\omega_0}$ & $\frac{\omega_0}{\alpha}$
& $\frac{T}{T_c}=\frac{m\omega_0^2k_BT}{\mu^2N}$ \\
 \hline
\end{tabular}
\end{center}
\caption{Nondimensional units and parameters.}
\label{t1}
\end{table}

It is convenient to render our equations dimensionless before we proceed with
their analysis. To do this, we can start with Eq. (\ref{2.3c}). The two terms in its left
hand side have the same order if we adopt $t^*=\omega_0 t$ as a nondimensional
time. The spins $\sigma_i$ are either +1 or -1, and therefore its right hand side (the forcing term) is, at most, $\mu N/m$. Adopting this value as an order of magnitude
of the forcing term, it is of the same order of magnitude as any of the terms
in the left side of (\ref{2.3c}) provided $x$ has an order of magnitude $[x]=
\mu N/(m\omega_0^2)$. The normalization condition
\begin{eqnarray}\label{2.14}
\sum_{\bm{\sigma}}\int_{-\infty}^\infty \rmd x \int_{-\infty}^\infty \rmd p\,
\mathcal{P}(x,p,\bm{\sigma},t)\,  =1,
\end{eqnarray}
yields
$$[\mathcal{P}] = \frac{1}{[x]\, [p]}= \frac{1}{m\omega_0[x]^2}=\frac{m\omega_0^3}{\mu^2N^2}.$$
Lastly, the argument of the coefficient in Eq. (\ref{2.5}) has order of magnitude
$$\frac{\mu\, [x]}{k_B T} = \frac{\mu^2 N}{m\omega_0^2 k_B T}= \frac{T_c}{T},$$
where
\begin{equation}\label{2.15}
T_c=\frac{\mu^2 N}{m\omega_0^2 k_B}
\end{equation}
is a critical temperature whose role we will unveil later in the paper.

Thus we can define nondimensional variables according to $x^*=x/[x]$,
$t^*=t/[t]$, \dots, where the units $[x]$, $[t]$, \ldots are as defined in Table \ref{t1}.
Inserting these nondimensional variables in Equations (\ref{2.3c}), (\ref{2.4}),
(\ref{2.5}) and (\ref{2.6}), and dropping the asterisks in the result (so as not to clutter our formulas), we obtain the following nondimensional equations
\begin{eqnarray}\label{2.16}
   && \frac{d^2 x}{dt^2}+x=\frac{1}{N} \sum_{i=1}^N \sigma_i \sigma_{i+1} \, ,\\
&& \sum_{i=1}^N \left[ W_i(R_i\bm{\sigma}|x,p) {\cal P}(x,p,R_i\bm{\sigma},t)-
W_i(\bm{\sigma}|x,p){\cal P}(x,p,\bm{\sigma},t) \right]\nonumber \\
 && \quad \quad = \epsilon\left[ \partial_t +p\, \partial_x + \left(\frac{1}{N}
 \sum_{i=1}^n \sigma_i \sigma_{i+1}-x\right) \partial_p \right]
 {\cal P}(x,p,\bm{\sigma},t),   \label{2.17}\\
&& W_i(\bm{\sigma}|x,p) = \frac{1}{2}-\frac{\gamma(x)}{4}\,
\sigma_i (\sigma_{i-1}+\sigma_{i+1}), \label{2.18}\\
&&\gamma(x)=\tanh\left(\frac{2x}{\theta}\right),\quad
\epsilon= \frac{\omega_0}{\alpha} .\label{2.18bis}
\end{eqnarray}
In nondimensional units, the equilibrium distributions (\ref{2.7}) and (\ref{2.11}) are
\begin{eqnarray}\label{2.19}
    {\cal P}_{\text{eq}}(x,p,\bm{\sigma})= \frac{1}{Z} \exp\left[-\frac{N}{\theta}\,
    {\cal H}(x,p,\bm{\sigma}) \right] ,\\
    {\cal H}(x,p,\bm{\sigma})=\frac{p^2+x^2}{2}-\frac{x}{N}\sum_{i=1}^N
    \sigma_i\sigma_{i+1}, \label{2.20}
    \end{eqnarray}
    and
    \begin{eqnarray}
    {\cal P}_{\text{eq}}(x,p)=\frac{1}{Z} \exp \left[-\frac{N}{\theta}\left(\frac{p^2}{2}
    +\mathcal{V}_{\text{eff}}(x)\right) \right],   \label{2.21}\\
    \mathcal{V}_{\text{eff}}(x)= \frac{x^2}{2}- \theta\left[\ln\cosh\left(\frac{x}{\theta}\right)+\ln 2\right],\label{2.22}
\end{eqnarray}
respectively.

\section{Equilibrium points and phase transition}
\label{s3}

The maxima of ${\cal P}_{\text{eq}}(x,p)$ determine the most likely position and momentum of the oscillator coupled to the Ising system, $(\widetilde{x}_{\text{eq}},\widetilde{p}_{\text{eq}})$, when the total system is at equilibrium. These most likely values will be called {\em macroscopic equilibrium values} following van Kampen's terminology \cite{vk92}. As $N\to\infty$, the equilibrium mean values of $x$ and $p$ coincide with $\widetilde{x}_{\text{eq}}$ and $\widetilde{p}_{\text{eq}}$, respectively, whereas the corresponding variances tend to zero. Similarly, the equilibrium average value of an smooth function $f(x,p)$ tends to its macroscopic value: $\langle f(x,p)\rangle\sim f(\widetilde{x}_{\text{eq}},\widetilde{p}_{\text{eq}})$ as $N\to\infty$. Thus a macroscopic quantity has negligible fluctuations in the limit of infinitely many oscillators. Let us now calculate $\widetilde{x}_{\text{eq}}$ and $\widetilde{p}_{\text{eq}}$ and the corresponding variances. First, $\widetilde{p}_{\text{eq}}=0$  and the oscillator is at rest in equilibrium, as expected. Second, the oscillator macroscopic equilibrium positions are given by the solutions of the equation
\begin{equation}\label{3.1}
    \left.\frac{\rmd {\cal V}_{\text{eff}}(x)}{\rmd x}\right|_{x=\widetilde{x}_{\text{eq}}}=0 \, ,
\end{equation}
i.e.,
\begin{equation}\label{3.2}
  \widetilde{x}_{\text{eq}}- \tanh \left(\frac{\widetilde{x}_{\text{eq}}}{\theta}\right)=0 \, .
\end{equation}
Clearly $\widetilde{x}_{\text{eq}}=0$ is always a solution for any value of $\theta$. It
is the only solution for $\theta>1$, it corresponds to a maximum of ${\cal P}_{
\text{eq}}$ and is therefore stable. At $\theta=1$ two new stable equilibria issue from $\widetilde{x}_{\text{eq}}=0$ and exist for $\theta<1$. Note that in dimensional units, $\theta=\theta_c=1$ corresponds to $T=T_c$, the critical temperature defined in (\ref{2.15}). Besides, $T_c$ should be independent of $N$ in the large $N$ limit. This gives the scaling of $\mu$ with $N$ mentioned in the Introduction,
\begin{equation}\label{3.2b}
    \mu=\frac{\mu_0}{\sqrt{N}} \,,
\end{equation}
where $\mu_0$ is independent of $N$. Therefore,
\begin{equation}\label{3.2c}
    T_c=\frac{\mu_0^2}{m\omega_0^2 k_B} \, ,
\end{equation}
making use of (\ref{2.15}).

 As $\theta\rightarrow 1^-$, we find
\begin{equation}\label{3.3}
    \widetilde{x}_{\text{eq}}\sim\pm \sqrt{3 \left(1-\theta\right)} \, ,
\end{equation}
i.e. the usual scaling at pitchfork bifurcations, $\widetilde{x}_{\text{eq}}\propto
|\theta-1|^{1/2}$. The effective potential (\ref{2.22}) is continuous at $\theta=1$,
\begin{equation}\label{3.4}
    \widetilde{\cal V}_{\text{eff}}^{\text{eq}}+ \theta \ln 2\sim \frac{1}{2}\left(1-\frac{1}{\theta}\right) \widetilde{x}^{2}_{\text{eq}}=-\frac{3 \theta}{2} \left(\frac{1}{\theta}-1\right)^2,
\end{equation}
as $\theta\to 1^-$. Then the derivative of $\widetilde{\cal V}_{\text{eff}}^{\text{eq}}$ with respect to $T$ is also continuous at $\theta =1$. We have found a second order, continuous,  phase transition with classical critical exponents. The equilibrium
position $x$ can then be considered the order parameter of the transition: its macroscopic value vanishes for $\theta >1$ and is non-zero for $\theta <1$.

Why does this second order transition appear? At first, it seems surprising to find it in a 1d model with short-ranged interactions. In order to understand the physical reason for this behaviour, let us calculate the equilibrium probability ${\cal P}_{\text{eq}}(\bm{\sigma})$ of finding the spins in configuration $\sigma$, regardless of the values of $x$ and $p$. We shall integrate the probability density (\ref{2.19}), written as
\begin{equation}\label{3.5}
\fl   \quad {\cal P}_{\text{eq}}(x,p,\bm{\sigma})=\frac{1}{Z} \exp\left\{ -\frac{N}{\theta} \left[ \frac{p^2}{2}+\frac{1}{2} \left(x-\varphi(\bm{\sigma})\right)^2- \frac{\varphi(\bm{\sigma})^2}{2} \right] \right\} \, ,
\end{equation}
where
\begin{equation}\label{3.6}
    \varphi(\bm{\sigma})=\frac{1}{N}\,\sum_i \sigma_i \sigma_{i+1} \,
\end{equation}
over $x$ and $p$, with the result
\begin{equation}\label{3.7}
    {\cal P}_{\text{eq}}(\bm{\sigma})=\frac{1}{Z_{\text{mf}}} \exp \left[\frac{N}{2\theta} \varphi(\bm{\sigma})^2 \right] =\frac{1}{Z_{\text{mf}}} \exp \left[\frac{1}{2\theta N} \sum_{i,j}s_i s_j \right] \, .
\end{equation}
Here $s_i=\sigma_i \sigma_{i+1}$ are new effective spin variables and $Z_{\text{mf}}$ is the appropriate normalization constant. Then (\ref{3.7}) corresponds to the
equilibrium probability of a mean field Ising model. Each spin $s_i$ is coupled to the global mean field $\varphi=\sum_j s_j/N$ and an effective long range interaction appears in the model. It is a well-known result that the 1d mean field Ising model has a second order phase transition at a finite temperature \cite{Re98}. The macroscopic, most probable, value of $\varphi$ is given by solutions of the trascendental equation \cite{Re98,BW34}
\begin{equation}\label{3.8}
    \widetilde{\varphi}_{\text{eq}}=\tanh \left( \frac{ \widetilde{\varphi}_{\text{eq}}}
    {\theta} \right) \, .
\end{equation}
There appears a second order transition at a critical temperature $\theta=1$, which is the same one appearing in Eq. (\ref{3.3}). The origin of this transition is the {\em hidden} long-range effective coupling between spins (\ref{3.7}) which is produced by the coupling of the Glauber spins to the oscillator. Similar hidden long-range effective
correlations that enable possible 1d phase transitions are present in biophysical
systems. An example is DNA melting \cite{Pe06} which has been modeled
by means of modified Ising systems \cite{DyP93,WyB85}, different from the one considered here. For DNA melting, the order of the phase transition has not
yet been well established: depending on models and conditions, it has been
predicted to be first order \cite{KMyP00}, second order \cite{HOyM08}, or even
higher \cite{GyT06}.

The fluctuations of the order parameter $x$ can be analyzed from the equilibrium distribution (\ref{2.21}) in the limit $N\to\infty$. The average of any function of
$x$ can be calculated by using the Laplace method in integrals involving
(\ref{2.21}), which leads to expanding the effective potential (\ref{2.22}) around the macroscopic value $\widetilde{x}_{\text{eq}}$. The result is
\begin{equation}\label{3.9}
  N\left[{\cal V}_{\text{eff}}(x)-\widetilde{{\cal V}}_{\text{eff}}^{\text{eq}}\right]=\frac{N}{2} \omega^2 \left( x-\widetilde{x}_{\text{eq}} \right)^2+\Or \left(N ( x-\widetilde{x}_{\text{eq}})^3 \right) \, ,
\end{equation}
where
\begin{equation}\label{3.10}
    \omega^2=1-\frac{1}{\theta}\left( 1- \widetilde{x}_{\text{eq}}^{2}\right) .
\end{equation}
$\omega$ is a new dimensionless frequency. Therefore, for $\omega\neq 0$ ($\theta\neq 1$), the fluctuations of $x$ are Gaussian because higher order terms vanish as $N\rightarrow\infty$. The average value of $x$ equals $\widetilde{x}_{\text{eq}}$, as expected, and its variance is
\begin{equation}\label{3.11}
   \sigma_x^2 \equiv \langle \left( x- \widetilde{x}_{\text{eq}} \right)^2  \rangle_{\text{eq}}=\frac{\theta}{N\omega^2}=\frac{\theta}{N} \left[1-\frac{1}{\theta}\left( 1-\widetilde{x}_{\text{eq}}^{2} \right)\right]^{-1} \, ,
\end{equation}
which vanishes as $N^{-1}$. Similarly, $\sigma_p^2\equiv\langle p^2\rangle_{\text{eq}} =\theta/N$. Since $\widetilde{x}_{\text{eq}}$ is of order one for $N\gg 1$ and $\theta\neq 1$, the fluctuation of $x$ is much smaller than its average value, which is the expected behaviour of a macroscopic variable. The average value of ${\cal V}_{\text{eff}}$ at equilibrium verifies
\begin{equation}
    N\left[\langle {\cal V}_{\text{eff}}\rangle_{\text{eq}}-\widetilde{{\cal V}}_{\text{eff}}^{\text{eq}}\right]=\frac{N}{2}\omega^2 \langle \left( x-\widetilde{x}_{\text{eq}} \right)^2\rangle =\frac{\theta}{2} \,, \label{3.12}
\end{equation}
for $\theta\neq 1$. The term coming from Gaussian fluctuations is subdominant in the thermodynamic limit as compared to the extensive macroscopic contribution $N\widetilde{{\cal V}}_{\text{eff}}^{\text{eq}}$.

On the other hand, $\omega\rightarrow 0$ and therefore $\sigma_x$ in (\ref{3.11})
diverges as $\theta\to 1$: fluctuation divergence is connected to the vanishing of the renormalized frequency $\omega$. A very large value of $N$, diverging for $\theta\to 1$, has to be considered in order to be in the ``thermodynamic limit'' for the oscillator position, where $x$ is approximately equal to its most probable value $\widetilde{x}_{\text{eq}}$ and its fluctuactions can be neglected. What happens for $\theta= 1$?
The first three differentials of the effective potential vanish at $\theta=1$ whereas
$d^4\mathcal{V}_{\text{eff}}/dx^4=2/\theta^3$. Then, as $\theta\to 1^+$
\begin{equation}\label{3.13}
N\,\frac{\mathcal{V}_{\text{eff}}-\widetilde{\mathcal{V}}_{\text{eff}}^{\text{eq}}}{\theta} \sim
\frac{N\, (\theta-1)}{2}\, (x-\widetilde{x}_{\text{eq}})^2+\frac{N (x-\widetilde{x}_{\text{eq}})^4}{12},
\end{equation}
and a similar expression holds as $\theta\to 1^-$ (replacing $(\theta-1)$ in
(\ref{3.13}) by $2(1-\theta)$). Therefore the fluctuations scale is $(x-\widetilde{x}_{\text{eq}})\propto N^{-1/4}$ as $N\to\infty$ if $|\theta-1|\ll N^{-1/2}\ll 1$
(non-Gaussian behaviour, the quadratic term can be neglected in comparison to the
quartic term) and as $(x-\widetilde{x}_{\text{eq}})\propto N^{-1/2}$ if $N^{-1/2}\ll
|\theta-1|$ (Gaussian behaviour, the quartic term is negligible).

\section{Dynamics}
\label{s4}

In this section we shall analyze the dynamical equations of motion. Equation  (\ref{2.16}) is a stochastic differential equation for $x$ because the configuration of the spin system $\bm{\sigma}$ is a stochastic process. Let us denote $C_{i,n}\equiv \sigma_i\sigma_{i+n}$, with $i=1,\ldots,N$ and $n\geq 0$. Of course,  $C_{i,0}=\sigma_i^2=1 $ for all $i$. By averaging (\ref{2.16}) over the joint probability ${\cal P}(x,p,\bm{\sigma},t)$ solution of the Liouville-master equation (\ref{2.6}), we obtain
\begin{equation}\label{4.-1}
   \frac{\rmd^2 \langle x\rangle}{\rmd t^2}+ \langle x\rangle =\frac{1}{N}\sum_{i=1} \langle C_{i,1}\rangle \, .
\end{equation}
From the Liouville-master equation, we can derive the following system of equations for the spin correlations
\begin{equation}\label{4.2}
\fl \qquad -2 \langle C_{i,n} \rangle +\frac{1}{2} \langle \gamma(x) \left( C_{i,n-1}+C_{i,n+1}+ C_{i-1,n+1}+C_{i+1,n-1} \right) \rangle = \epsilon\, \frac{\rmd}{\rmd t}\langle C_{i,n} \rangle\,,
\end{equation}
for $n\geq 1$ and $i=1,\ldots, N$. Here $\gamma(x)$ and $\epsilon$ are given by (\ref{2.18bis}). The system of equations (\ref{4.2}) must be solved with the boundary condition $C_{i,0}=1$ and given initial conditions $\{\langle C_{i,n} \rangle(t=0);n\geq 1\}$.

As explained in section \ref{s3}, a quantity is called macroscopic if, compared to its mean, its fluctuations are negligible in the limit as $N\to\infty$. In this section, we will describe the mean-field (macroscopic) dynamics of our oscillator-spin system such that
\begin{eqnarray}
\langle F(x,p,\sigma_i\sigma_{i+n})\rangle(t)&=&\int \rmd x \int \rmd p \sum_{\bm{\sigma}}
F(x,p,\sigma_i \sigma_{i+n})\, {\cal P}(x,p,\bm{\sigma},t)\nonumber\\
& \sim& F(\widetilde{x}(t),\widetilde{p}(t),\widetilde{C_{i,n}}(t)), \label{4.0}\\
\qquad\langle\sigma_i(t)\sigma_{i+n}(t)\rangle &=&\langle C_{i,n}\rangle(t)\sim \widetilde{C_{i,n}}(t),\quad i=1, \ldots, N, \label{4.4}
\end{eqnarray}
in the limit as $N\to\infty$ for any smooth function $F(x,y,z)$. In these equations and for each time $t$, $\widetilde{x}(t)$, $\widetilde{p}(t)$ and $\widetilde{C_{i,n}}(t)$ are the values of $x$, $p$ and of $C_{i,n}$ for which the probability density function ${\cal P}(x,p,\bm{\sigma},t)$ has a maximum.  Due to translation invariance, in the limit as $N\to\infty$, the averages $\langle C_{i,n}\rangle(t)$ are independent of $i$, provided the initial probability density is translation invariant and isotropic. Then $\widetilde{C_{i,n}}(t)$ is independent of $i$ and we can write $\widetilde{C_n}(t)$ instead of $\widetilde{C_{i,n}}(t)$ (or $\langle C_{i,n}\rangle(t)$) in (\ref{4.-1})-(\ref{4.4}). Since $\sigma_i^2=1$, we have $\widetilde{C}_0=1$. Ignoring fluctuations according to (\ref{4.0}), equation (\ref{4.-1}) yields
\begin{equation}\label{4.1}
  \frac{\rmd^2\widetilde{x}}{\rmd t^2}+ \widetilde{x} =\widetilde{C_1} \, ,
\end{equation}
and (\ref{4.2}) simplifies to
\begin{equation}\label{4.6}
  -2 \widetilde{C_n}+ \gamma(\widetilde{x}) \left( \widetilde{C_{n-1}}+\widetilde{C_{n+1}} \right)=  \epsilon\,\frac{\rmd}{\rmd t}\widetilde{C_n}\, , \qquad n\geq 1 \, ,
\end{equation}
corresponding to van Kampen's macroscopic approximation \cite{vk92}. This approximation is equivalent to separating macroscopic and fluctuating contributions in $x$ and $C_{i,n}$:
\begin{eqnarray}
  x &=& \widetilde{x}+ \delta x  \, ,
  \label{4.5a}
  \\
  C_{i,n} &=& \widetilde{C_n}+ \delta C_{i,n} \, .
    \label{4.5b}
\end{eqnarray}
Inserting these expressions in (\ref{4.2}) and neglecting all terms containing
correlations, such as $\langle(\delta x)^2\rangle$ or $\langle\delta x \delta C_{i,n}\rangle$, we obtain again (\ref{4.6}).
The mean-field or macroscopic dynamical behaviour of the oscillator-spin system is found by solving the equations (\ref{4.1}) and (\ref{4.6}) with the boundary condition $\widetilde{C_0}=1$ and appropriate initial conditions.

We now consider the limit $\epsilon=\omega_0/\alpha\ll 1$ of a very slow oscillator compared to the relaxation time of the Glauber spins. Setting $\epsilon=0$ in (\ref{2.17}), we find the equilibrium solution of the master equation for each instantaneous value of $x(t)$. For $\epsilon\ll 1$, there is an initial time window inside which the $\widetilde{C_n}$ reach their equilibrium values (corresponding to the equilibrium solution of the master equation with fixed $x(t)$) while the oscillator position and velocity are frozen at their initial values. After this initial layer, we can approximately solve (\ref{4.2}) by means of an expansion in powers of $\epsilon$:
\begin{eqnarray}
\widetilde{C_n}(t;\epsilon)=\sum_{k=0}^{1}\widetilde{C_n^{(k)}}(t)\, \epsilon^k
+\Or(\epsilon^2). \label{4.7}
\end{eqnarray}
This yields the boundary conditions $\widetilde{C_0^{(0)}}=1$,
$\widetilde{C_0^{(1)}}=0$. Inserting (\ref{4.7}) into (\ref{4.6}), we obtain the
following system of equations
\begin{eqnarray}
&&\gamma(\widetilde{x})(\widetilde{C^{(0)}_{n-1}}+
\widetilde{C^{(0)}_{n+1}})-2 \widetilde{C^{(0)}_n}= 0,\label{4.8}\\
&&\gamma(\widetilde{x})(\widetilde{C^{(1)}_{n-1}}+
\widetilde{C^{(1)}_{n+1}})-2 \widetilde{C^{(1)}_n}= \frac{\rmd\widetilde{C^{(0)}_n}}{
\rmd t},\label{4.9}
\end{eqnarray}
and so on. The solutions of (\ref{4.8}) and (\ref{4.9}) with boundary conditions
$\widetilde{C^{(0)}_0}=1$ and $\widetilde{C^{(1)}_0}=0$ are found in
\ref{apa}. They provide
\begin{equation}\label{4.10}
\widetilde{C_1}=\tanh\left(\frac{\widetilde{x}}{\theta}\right)-\frac{\epsilon}{2\theta}
\frac{1+\tanh^2\left(\frac{\widetilde{x}}{\theta}\right)}{1-\tanh^2\left(\frac{\widetilde{x}
}{\theta}\right)}\, \frac{\rmd\widetilde{x}}{\rmd t}\,.
\end{equation}
Eq. (\ref{4.10}) comes from a ``normal'' solution of the system of equations \cite{ByP93,ByP94}, in which all the time dependence in $\widetilde{C_n}$ occurs through $\widetilde{x}$, which evolves on the slower time scale $t$. The first term in (\ref{4.10}) is the equilibrium value of $\widetilde{C_1}$ corresponding to the instantaneous oscillator position $\widetilde{x}$, whereas the second term contains the (small) deviations from equilibrium, to order $\epsilon\ll 1$. The initial condition $\widetilde{C_1}(t=0)$ does not appear in expression (\ref{4.10}) because the spins forget their initial conditions on a time scale (initial layer) much shorter than the natural period of the oscillator. Inserting (\ref{4.10}) into (\ref{4.1}), we get
\begin{equation}\label{4.11}
\frac{\rmd^2\widetilde{x}}{\rmd t^{2}}+ \frac{\epsilon}{2\theta} \frac{1+\tanh^2(
\frac{\widetilde{x}}{\theta})}{1-\tanh^2(\frac{\widetilde{x}}{\theta})}\, \frac{\rmd
\widetilde{x}}{\rmd t}+\widetilde{x}- \tanh\left(\frac{\widetilde{x}}{\theta}\right)=0.
\end{equation}

Equation (\ref{4.11}) can be rewritten in terms of the nondimensional
effective potential (\ref{2.22}) and the friction coefficient
\begin{equation}\label{4.12}
R(\widetilde{x}) =\frac{1+\tanh^2( \frac{\widetilde{x}}{\theta})}{1-\tanh^2( \frac{\widetilde{x}}{\theta})} ,
\end{equation}
as
\begin{equation}\label{4.13}
\frac{\rmd^2\widetilde{x}}{\rmd t^{2}}=- \mathcal{V}^{\prime}_{\text{eff}}(
\widetilde{x})- \frac{\epsilon}{2\theta}R(\widetilde{x})\, \frac{\rmd\widetilde{x}}{\rmd t}.
\end{equation}
The equilibrium values of $x$ can be obtained from  (\ref{4.11}) and the results
of the previous section are recovered. Again stable equilibrium points correspond to the minima of ${\cal V}_{\text{eff}}$.

Equation (\ref{4.11}), or equivalently (\ref{4.13}), is the main result of this section.
It shows that the effect of the coupling of the oscillator with the bath of Ising spins is
twofold. Firstly, the potential is ``renormalized'' to ${\cal V}_{\text{eff}}$, as a new force $\tanh(\widetilde{x}/\theta)$ is added to the harmonic interaction $-\widetilde{x}$; secondly, a nonlinear friction term proportional to $\rmd\widetilde{x}/\rmd t$ appears.

Which is the expected range of validity of the nonlinear equation (\ref{4.11})? For a given value of $\epsilon$, the range depends on the order of $\widetilde{x}$ and $\theta$. Equation (\ref{4.11}) holds if the spins relax to equilibrium so fast that the oscillator position does not change. Using that $\lambda_m=2 [1-\gamma(\widetilde{x})]$ is the smallest eigenvalue of the coefficient matrix in (\ref{4.6}) \cite{Gl63}, we find $\lambda_m\sim 2$ for $\theta\gg\widetilde{x}$ (high temperature limit) and $\lambda_m\sim\epsilon$ for
\begin{equation}\label{4.13b}
    2 \left[ 1-\tanh\left(\frac{2 \widetilde{x}}{\theta}\right)\right] \sim \epsilon \Longrightarrow \frac{\widetilde{x}}{\theta}\sim -\frac{1}{4} \ln \left(\frac{\epsilon}{4}\right) \, .
\end{equation}
For $\widetilde{x}/\theta$ satisfying (\ref{4.13b}) and larger values, our separation of time scales breaks down and we do not expect (\ref{4.11}) to hold. Due to the logarithmic dependence on $\epsilon$ in (\ref{4.13b}), our asymptotic theory should already fail for moderate values of $\widetilde{x}/\theta$. For instance, $\widetilde{x}/\theta\approx 1.5$ for $\epsilon=0.01$. More will be said about this point in the numerical section. We now particularize our theory for temperatures above and below critical.

\subsection{The region $\theta>1$}
\label{s4a}

In the high temperature region $\theta>1$, the stable equilibrium point of the oscillator is $\widetilde{x}_{\text{eq}}=0$. This equilibrium is asymptotically stable due to the presence of the damping term, and therefore $\widetilde{x}(t)\ll 1$ for long enough times.  Then (\ref{4.11}) can be approximated as:
\begin{equation}\label{4.14}
    \frac{\rmd^2\widetilde{x}}{\rmd t^{2}}+\frac{\epsilon}{2\theta}
    \frac{\rmd\widetilde{x}}{\rmd t} + \left(1-\frac{1}{\theta}\right) \widetilde{x} =0 \, ,
\end{equation}
which is the equation of the damped harmonic oscillator with square frequency
$1-1/\theta$ (which equals the renormalized frequency
defined in (\ref{3.10}) for $\widetilde{x}_{\text{eq}}=0$) and friction coefficient $\epsilon/(2\theta)$.
The renormalized frequency tends to zero as $\theta\to 1^+$.

Defining the damping ratio
\begin{equation}\label{4.16}
    \zeta=\frac{\epsilon}{4\sqrt{\theta(\theta-1)}},
    \end{equation}
the underdamped, critically damped and overdamped oscillations correspond to $\zeta<1$, $\zeta=1$ and $\zeta>1$, respectively. Thus, a new dynamical ``critical'' temperature $\theta_d^+$ appears for $\theta>1$, defined by the condition $\zeta=1$. According to (\ref{4.16}), this occurs for
\begin{equation}\label{4.17}
    \theta_d^{+}=\frac{1}{2} \left( 1+\sqrt{1+\frac{\epsilon^2}{4} } \right) = 1+
    \frac{\epsilon^2}{16} +\Or(\epsilon^{4})\, .
\end{equation}
Then, in the limit $\epsilon=\omega_0/\alpha \ll 1$ we are analyzing, $\theta_d^+$ is very close to the critical temperature $1$ and the region of overdamped oscillations is very narrow: its width is of the order of $\epsilon^{2}$.

\subsection{The region $\theta<1$}
\label{s4b}

In this region, the equilibrium position of the oscillator is given by the nonvanishing  solutions of (\ref{3.8}), $\pm\widetilde{x}_{\text{eq}}$. Thus the dynamics will be governed by the nonlinear equations (\ref{4.11}). If $\widetilde{x}_1(t)$ is one  solution evolving towards $\widetilde{x}_{\text{eq}}$ as $t\to\infty$,
$-\widetilde{x}_1(t)$ is also a solution which evolves towards $-\widetilde{x}_{\text{eq}}$. This is not in contradiction with the {\em linear} Liouville-master equation having a unique equilibrium distribution (\ref{2.19})-(\ref{2.20}). In fact, the multiplicity of (macroscopic) equilibrium solutions corresponding to extrema of the equilibrium
distribution is a direct consequence of the nonlinearity of the macroscopic equation, which is compatible with the linearity of the master equation \cite{vk92}.

Near the stable equilibrium points,
\begin{equation}\label{4.18}
    \widetilde{x}=\widetilde{x}_{\text{eq}}+\xi \, ,
\end{equation}
with $\xi\ll 1$, we linearize (\ref{4.11}) or (\ref{4.13}), thereby obtaining
\begin{equation}\label{4.19}
    \frac{\rmd^2 \xi}{\rmd t^{2}}+r_-\frac{\rmd\xi}{\rmd t} +\omega^2_- \xi =0\, .
\end{equation}
Here,
\numparts
\begin{eqnarray}
    \omega_{-}^2 &=& \mathcal{V}^{\prime\prime}_{\text{eff}}(
    \widetilde{x}_{\text{eq}})= 1-\frac{1-\widetilde{x}^{2}_{\text{eq}}}{\theta} ,
    \label{4.20a} \\
    r_- & = & \frac{\epsilon}{2\theta}\,R(\widetilde{x}_{\text{eq}})=
    \frac{\epsilon}{2\theta}\,\frac{1+\widetilde{x}^{2}_{\text{eq}}}{1- \widetilde{x}^{2}_{\text{eq}}} . \label{4.20b}
\end{eqnarray}
The frequency $\omega_-$ is equal to the renormalized frequency $\omega$ introduced in (\ref{3.10}), particularized for $\theta<1$. As in the case $\theta>1$, we find the
equations of a damped harmonic oscillator, but with different friction coefficient and frequency. As $\theta\to 1^-$, we get
\begin{equation}\label{4.20c}
    \omega_-^2 \sim  2  \left(1- \theta \right) \, ,
\end{equation}
\endnumparts
which also tends to zero.

The analysis of the overdamped, critically damped and underdamped oscillation regions is completely analogous to the case $\theta>1$. We find a new temperature $\theta_d^-$, for which the oscillations are critically damped. Thus for $\theta<\theta_d^-$ the oscillations are underdamped, while for $\theta_d^-<\theta<1$ they are overdamped. The critical dynamical temperature $\theta_d^{-}$ is determined by
\begin{equation}\label{4.21}
    \frac{r_-}{2\omega_-}=1 \, ,
\end{equation}
which gives, after some calculation,
\begin{equation}\label{4.22}
    \theta_d^- = 1-\frac{\epsilon^2}{32}+\Or(\epsilon^4)\, .
\end{equation}
Again, the region of overdamped oscillations below $\theta=1$ is very narrow.
It should be noted that another region of overdamped oscillations is predicted by (\ref{4.19})-(\ref{4.20b}) for very low temperatures, as the friction coefficient $r_-$ formally diverges for $T\to 0$ or $\widetilde{x}_{\text{eq}}\to 1$. Nevertheless, we will not investigate this region because it lies outside the range of validity of our dynamical equation (\ref{4.11}), as we discuss in relation with the numerical results in the next section.

\section{Numerical results}
\label{s5}

In order to test our theoretical predictions, we have carried out numerical simulations of the stochastic process corresponding to the dimensionless Liouville-master equation (\ref{2.17})-(\ref{2.18}). Equivalently, we have to integrate numerically the oscillator equation (\ref{2.16}) and the Glauber evolution equations for the spins given by the transition probabilities (\ref{2.18}). Using
the initial probability distribution ${\cal P}(x,p,\bm{\sigma},0)$, we generate initial conditions $(x(0;\nu),p(0;\nu),\bm{\sigma}(0;\nu))$ for $N_T$ trajectories $\nu$ ($\nu=1,\ldots,N_T$). The oscillator position and momentum and the spin configuration of a given trajectory $\nu$ at time $t$ are denoted by $x(t;\nu)$, $p(t;\nu)$ and $\bm{\sigma}(t;\nu)$, respectively. For a given trajectory at time $t$, we choose at random one spin $\sigma_i$ and flip it with probability $W_i(\bm{\sigma}|x(t;\nu),p(t;\nu))\leq 1$ at time $t+\Delta t$ (in our dimensionless time scale with time unit $1/\omega_0$, $\Delta t=\epsilon/N$, according to the Metropolis algorithm for the master equation \cite{MRRTyT53,NyB99}; in dimensional units, we have $\Delta t=\epsilon/(N\omega_0)=(N\alpha)^{-1}$  ). The oscillator position and momentum are also updated by
\begin{eqnarray}
  x\left(t+\frac{\epsilon}{N};\nu\right) &=& x(t;\nu)+\frac{\epsilon}{N}\, p(t;\nu) \, , \label{5.1a} \\
  p\left(t+\frac{\epsilon}{N};\nu\right) &=& p(t;\nu)+\frac{\epsilon}{N} \left[ -x(t;\nu)+
  \frac{1}{N} \sum_i \sigma_i(t;\nu) \sigma_{i+1}(t;\nu) \right] \, .
\end{eqnarray}
Up to a certain time $t_0$, each trajectory $\nu$ is obtained by iterating this procedure $t_0 N \epsilon^{-1}$ times. Afterwards, the numerical averages over $\nu$ (trajectories) give the averages with the probability distribution ${\cal P}(x,p,\bm{\sigma})$. In particular, we get the self-averaging properties,
\begin{eqnarray}
\langle x\rangle (t) &=& \frac{1}{N_T}\sum_{\nu=1}^{N_T} x(t;\nu)\to \tilde{x}(t), \label{5.2a} \\
\langle C_1(t)\rangle &=& \frac{1}{N N_T} \sum_{\nu=1}^{N_T}\sum_{i=1}^N \sigma_i(t;\nu) \sigma_{i+1}(t;\nu)\to\tilde{C_1}(t),\label{5.2b}
\end{eqnarray}
as $N\to\infty$ and $N_T\to\infty$. The simulation results $\langle x(t)\rangle$ and $\langle C_1(t) \rangle$ should therefore approach the macroscopic values $\tilde{x}(t)$ and $\tilde{C_1}(t)$, respectively, for sufficiently large $N$ and $N_T$.  The numbers of spins $N$ and of trajectories $N_T$ needed to get good approximations to $\widetilde{x}$ and $\widetilde{C_1}$ in (\ref{5.2a}) and (\ref{5.2b}) are related to the amplitude of the averaged trajectories. As this amplitude decreases, the number of particles and trajectories must be increased. When $\widetilde{x}(0)=\Or(1)$ and  $\dot{\widetilde{x}}(0)=\Or(1)$, good averages are obtained with $N\geq 10^3$ and $N_T\geq 10^2$. We have used $N=10^4$ and $N_T=10^2$ in our numerical simulations although we have observed that, depending on the initial values of $x$ and $p$, we can take smaller $N$ and $N_T$ without losing accuracy.
Note that an order-one initial dimensionless position $\widetilde{x}$ of the oscillator corresponds to a dimensional position $x$ of order $\sqrt N$. This means that the oscillator energy is comparable to the energy of the spin system, which is also of order $N$.

Our theory is expected to provide a  good description of the numerical curves if $\epsilon\ll 1$ and $x(0)/\theta$ is not too large (high temperature). As the temperature decreases for fixed $x(0)$, the characteristic relaxation time of the Ising system increases and becomes comparable to the oscillator period when $x(0)/\theta$ satisfies (\ref{4.13b}). For lower temperature, we expect our theory to break down. Let us check this from the results of the numerical simulations.

\begin{figure}[htbp]
\begin{center}
\includegraphics[width=9cm]{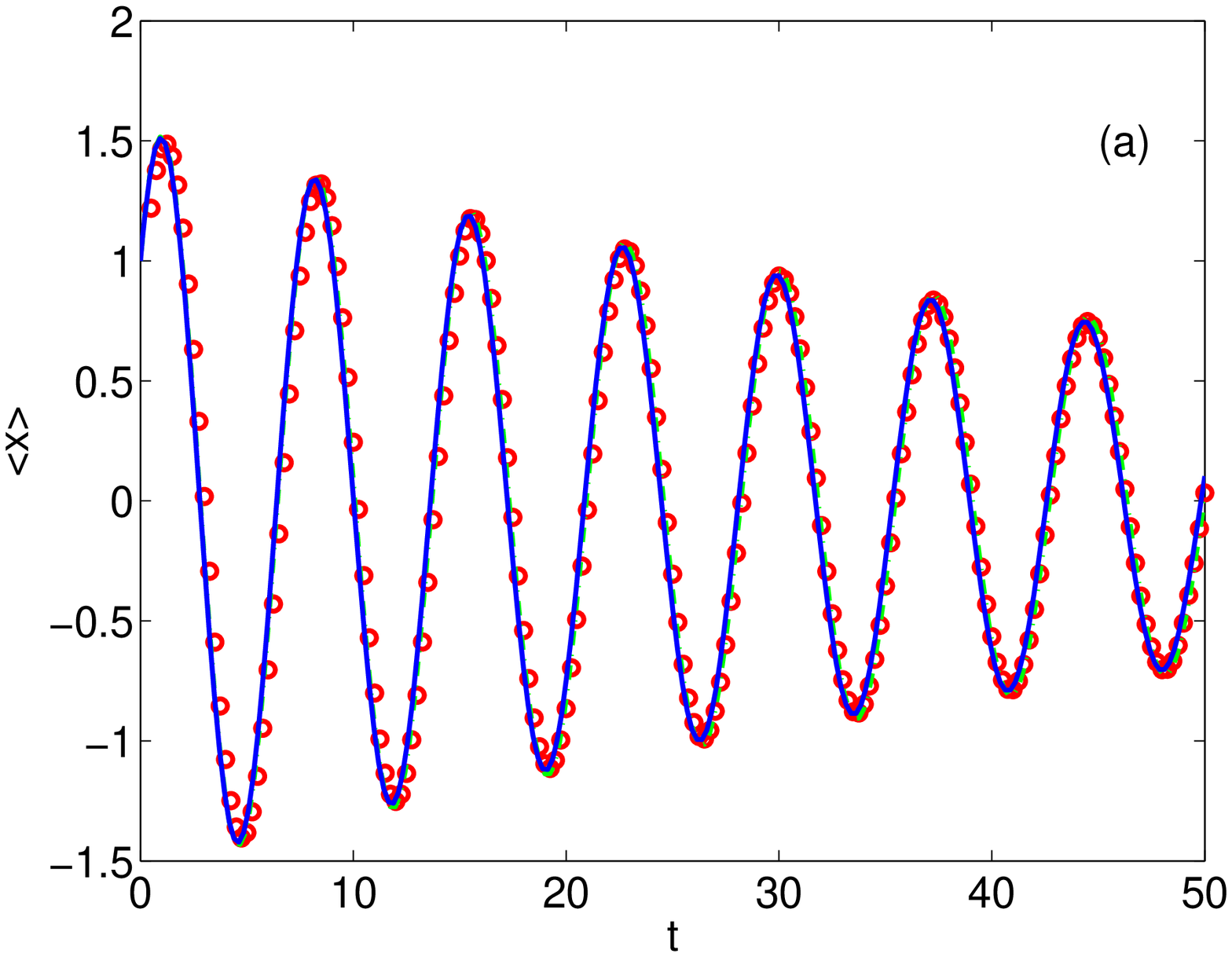}
\includegraphics[width=9cm]{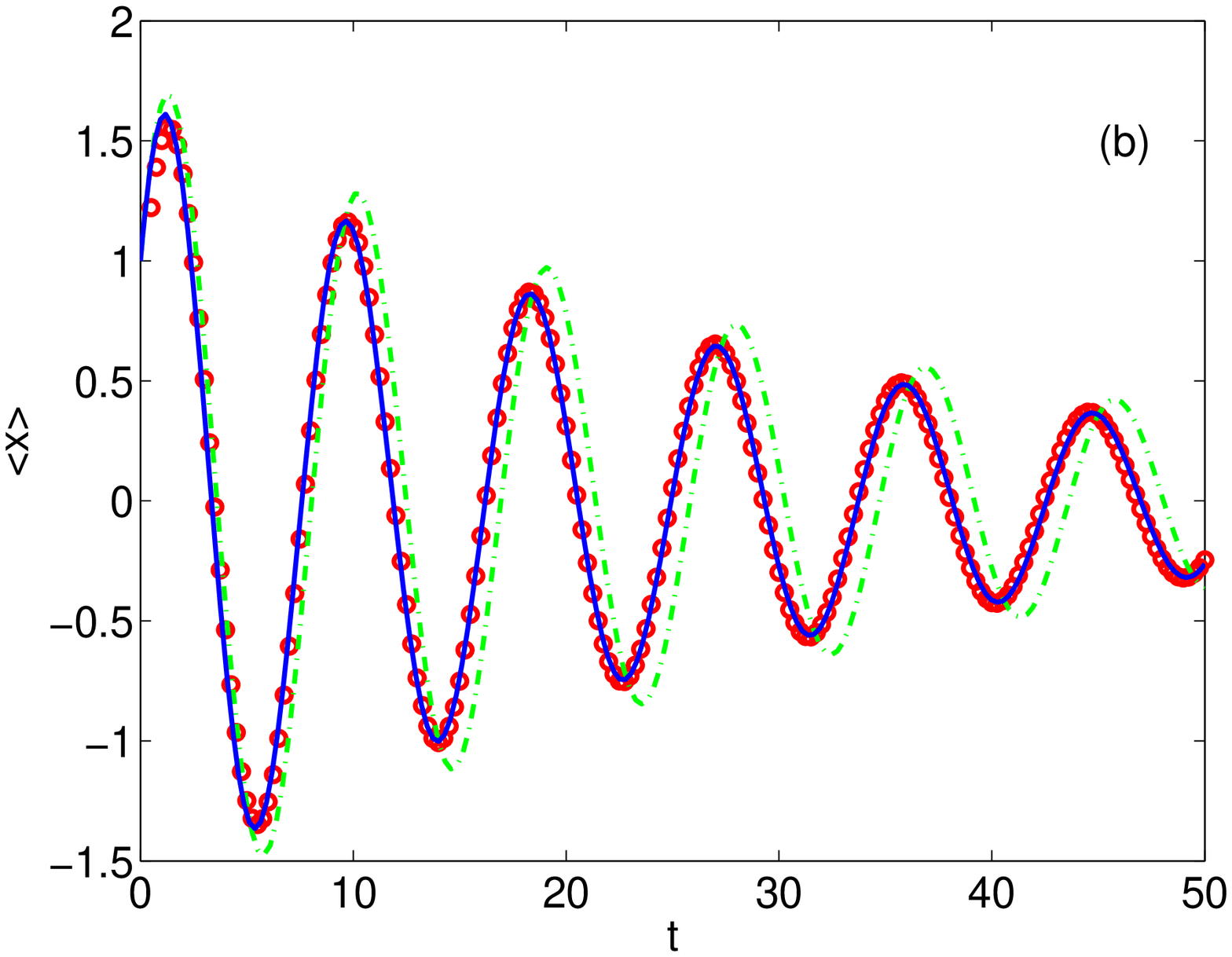}
\caption{Averaged trajectories $\langle x(t)\rangle=\widetilde{x}(t)$ (circles) versus
nonlinear (solid blue line) and linear (dot-dashed green line) predictions for
initial data $\widetilde{x}(0)=1$, $\dot{\widetilde{x}}(0)=1$ and (a) $\theta=4$,
(b) $\theta=2$. Other parameter values are $\epsilon=0.25$,
$N=10^4$, $N_T=10^2$.}
\label{fig1}
\end{center}
\end{figure}

In figure \ref{fig1}(a), we show the time evolution of the oscillator for one
high value of the temperature, namely $\theta=4$, corresponding to the underdamped region. We have chosen initial conditions so that the spin system is initially in a completely random state (therefore $C_1(0)=0$) and, for the oscillator $(\widetilde{x}(0),\dot{\widetilde{x}}(0))=(1,1)$. The numerical curves have been obtained with $N=10^4$ spins and averaged over $N_T=10^2$ trajectories. The theoretical predictions based on both the nonlinear evolution equation (\ref{4.11}) and the linear approximation (\ref{4.14}) (which are almost indistinguishable) show excellent agreement with the simulations. For the plotted values, $\tanh (\widetilde{x}/\theta)\simeq \widetilde{x}/\theta$ thereby justifying the use of the linear approximation (\ref{4.14}). Similar behaviour is obtained for different values of $\widetilde{x}(0)$ and $\dot{\widetilde{x}}(0)$, provided $\widetilde{x}/\theta\ll 1$ for all times. For larger $\dot{x}(0)$, the nonlinear evolution equation (\ref{4.11}) describes well the dynamics but there is a initial time window for which $\widetilde{x}/\theta$ is not small and the linear approximation is valid. Since given sufficient time, $\widetilde{x}\rightarrow 0$ for any initial condition, the linear equation (\ref{4.14}) always provides a good approximation of the dynamics for long enough times. Similarly, for the same initial conditions as in Figure \ref{fig1}(a), Figure \ref{fig1}(b) shows that the nonlinear equation (\ref{4.11}) gives a good approximation of the simulations for a lower temperature $\theta=2$ but the linear equation (\ref{4.14}) does not: $\tanh (\widetilde{x}/\theta)\simeq \widetilde{x}/\theta$ no longer holds.

\begin{figure}[htbp]
\begin{minipage}[b]{0.45\linewidth}
\centering
\includegraphics[width=\linewidth]{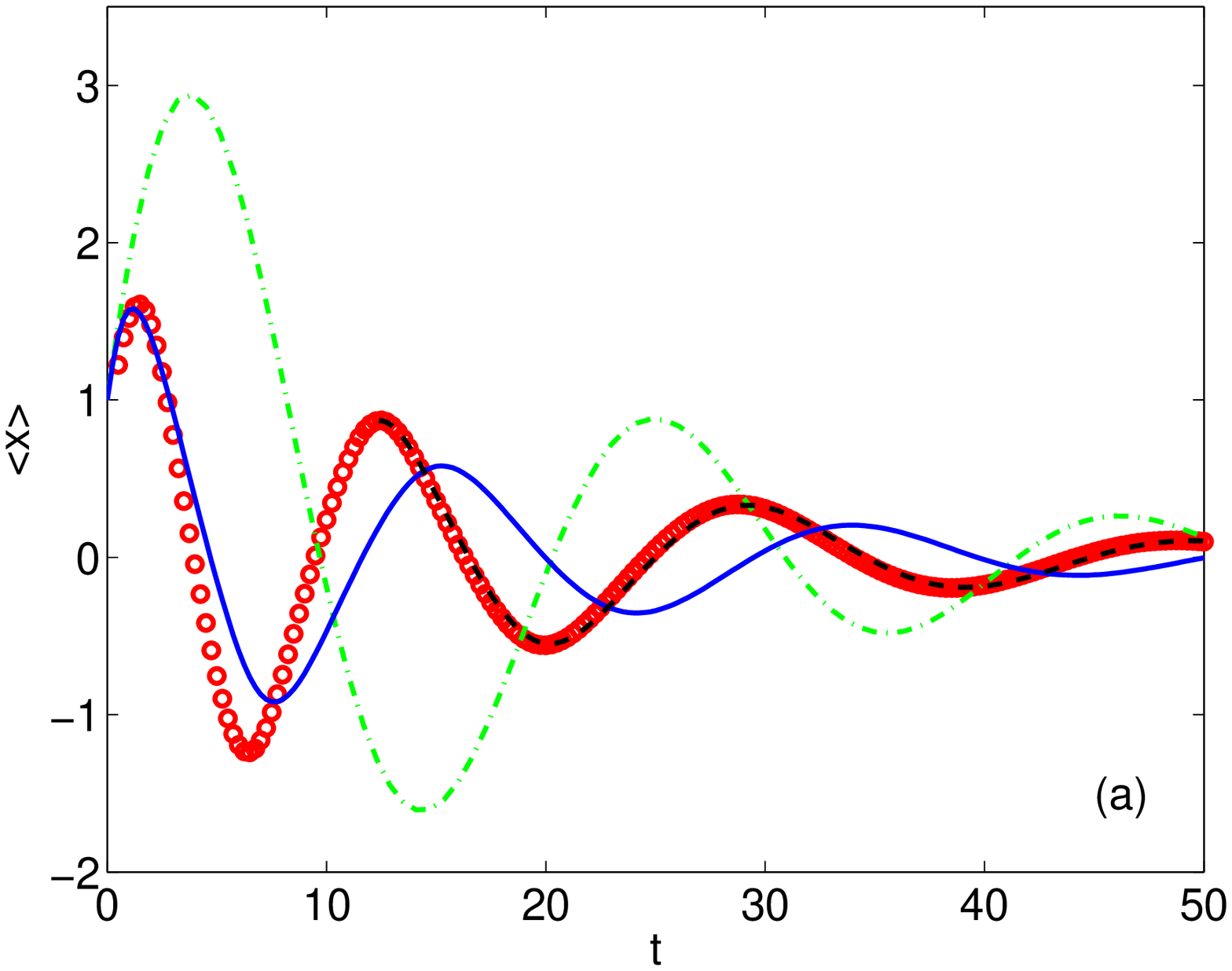}
\end{minipage}
\begin{minipage}[b]{0.45\linewidth}
\centering
\includegraphics[width=\linewidth]{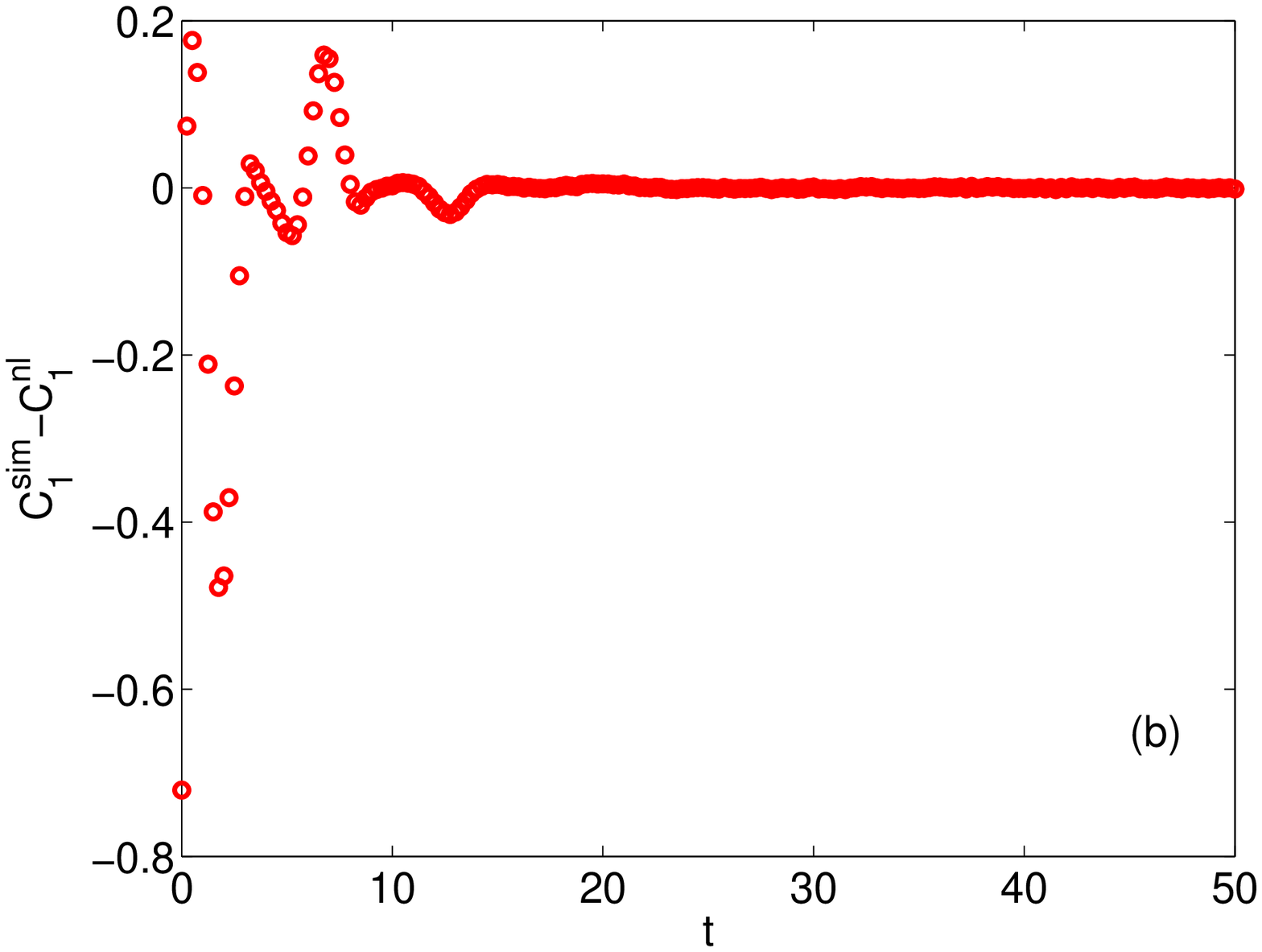}
\end{minipage}
\caption{(a) Comparison between the averaged trajectories (red circles), the nonlinear equation (\ref{4.11}) (solid blue line) and the linear equation (\ref{4.14}) (dashed green line) for $\theta=1.1$, near but above the transition temperature. (b) Difference between the average value $C_1^{\text{sim}}$ in the simulation and the theoretical prediction $C_1^{\text{nl}}$, (\ref{4.10}). }
\label{fig3}
\end{figure}

For $\theta=1.1$ (a value closer to the critical temperature $\theta=1$) and the same initial conditions, Figure \ref{fig3}(a) shows that even the predictions based on the nonlinear equation (\ref{4.11}) fail to approximate the simulation results. This could have been foreseen because Eq.\ (\ref{4.13b}) gives $\theta\simeq 1.44$ (for an initial $\widetilde{x}\simeq 1$) as the limiting temperature above which the nonlinear equation holds. Figure \ref{fig3}(b) shows that the difference between the average value $C_1^{\text{sim}}$ in the simulation and the theoretical prediction $C_1^{\text{nl}}$, (\ref{4.10}), is rather larger than the theoretical error of order $\epsilon^2=0.0625$ for times $t\leq 10$.

\begin{figure}[htbp]
\begin{center}
\includegraphics[width=9cm]{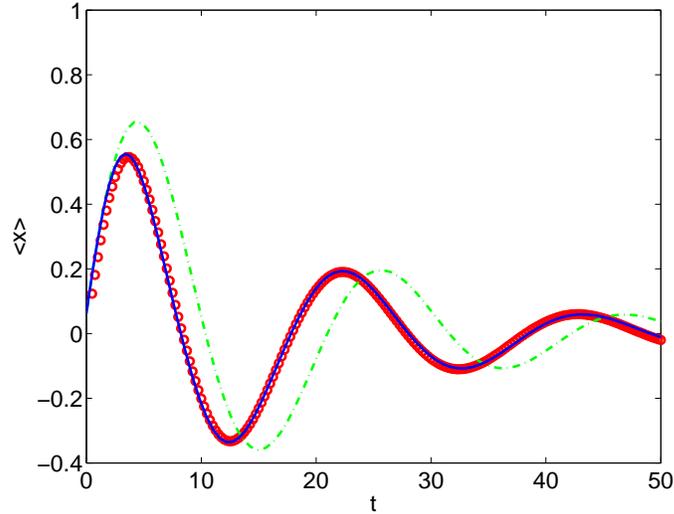}
\caption{Averaged trajectories $\langle x(t)\rangle=\widetilde{x}(t)$ (circles) versus
nonlinear (solid blue line) and linear (dot-dashed green line) predictions for
initial data $\widetilde{x}(0)=\epsilon^2$, $\dot{\widetilde{x}}(0)=\epsilon$ and $\theta=1.1$. Other parameter values are as in Figure \ref{fig1}.  }
\label{fig4}
\end{center}
\end{figure}

\begin{figure}[htbp]
\begin{center}
\includegraphics[width=6cm]{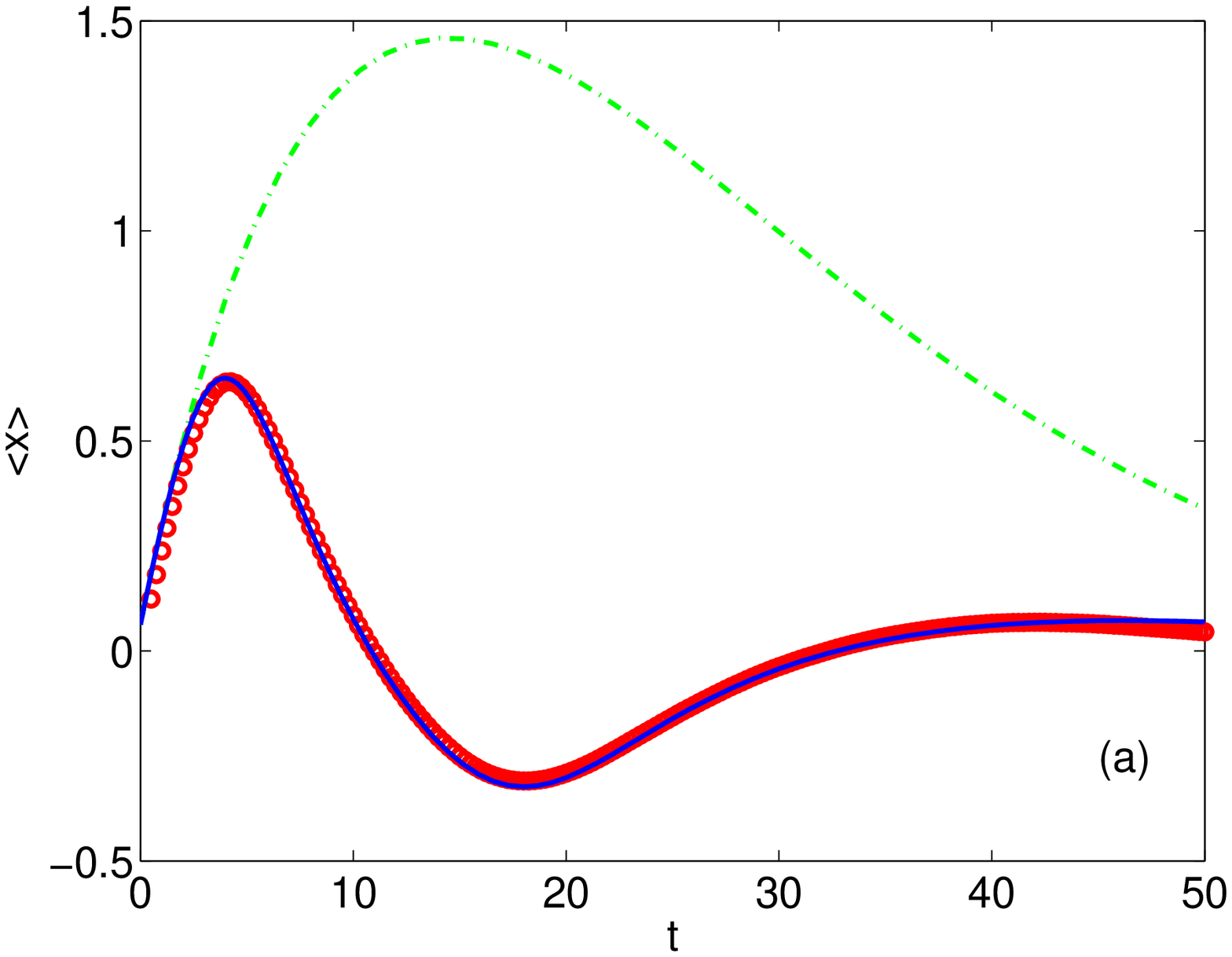}
\includegraphics[width=6cm]{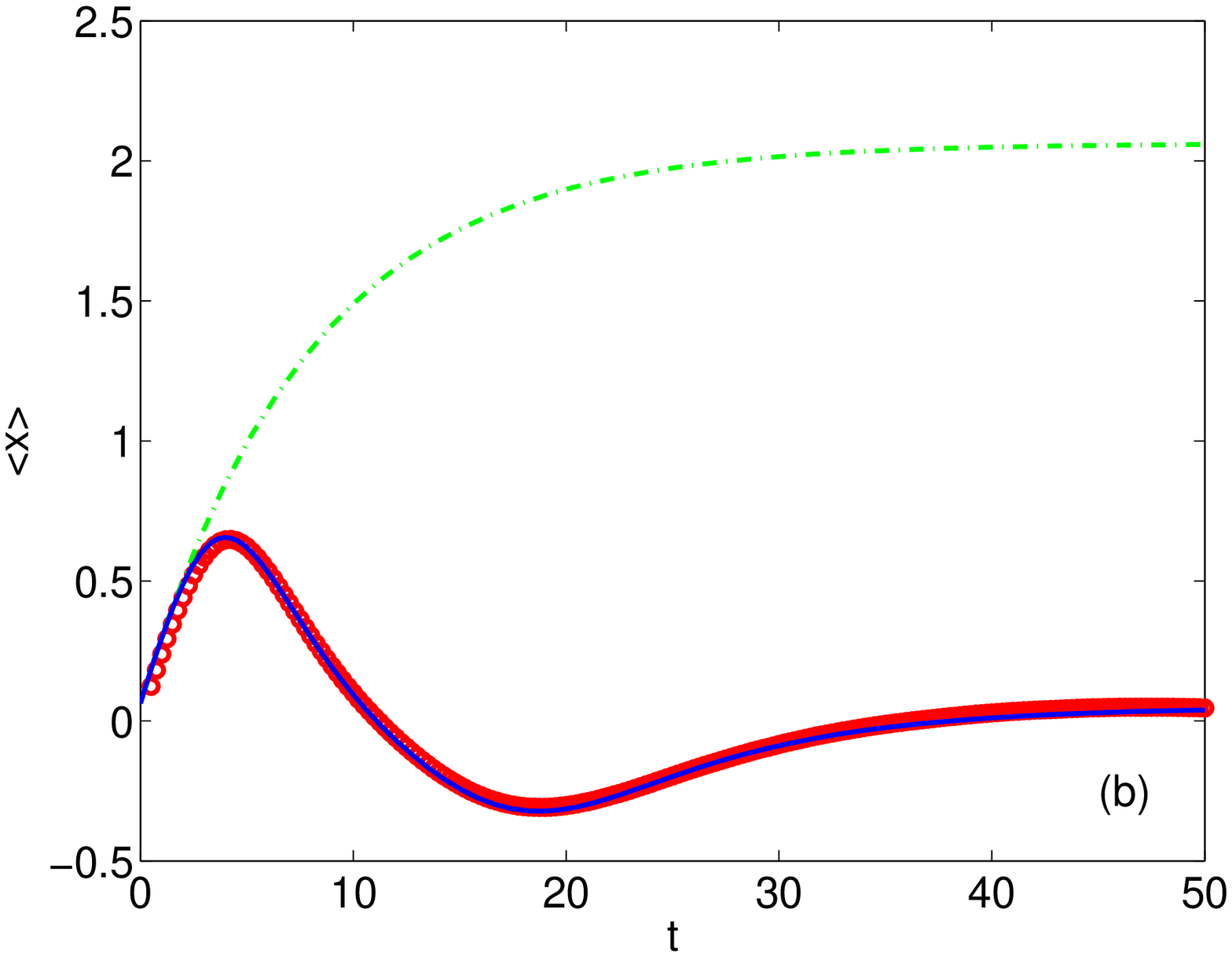}
\caption{Same as in Figure \ref{fig4} but with
(a) $\theta=1.005$, and (b) $\theta=1$.  }
\label{fig2}
\end{center}
\end{figure}

Selecting again the completely random state ($C_1^{\text{sim}}(0)$=0) as initial condition of the Ising spins, we have considered smaller values of $\widetilde{x}(0)$ and $\dot{\widetilde{x}}(0)$. They are such that initially the rhs of (\ref{4.10}) is of order $\epsilon^2$ and $C_1^{\text{sim}}(0)-C_1^{\text{nl}}(0)=\Or(\epsilon^2)$. Therefore, (\ref{4.10}) is initially valid and the possible departure from (\ref{4.11}) cannot be a transient effect, due to ``inadequate'' initial conditions. Equation (\ref{4.10}) suggests immediately the choice
\begin{equation}\label{5.3}
   \widetilde{x}(0)=\epsilon^2 \,, \quad \dot{\widetilde{x}}(0)=\epsilon \, .
\end{equation}
With this choice of initial conditions, the nonlinear equation (\ref{4.10}) is still a good approximation for the dynamical behaviour of the oscillator at temperature $\theta=1.1$, as shown by figure \ref{fig4}. Interestingly, this choice corresponds to variables of the order of unity for a different, alternative, nondimensionalization of the variables,
\begin{equation}\label{5.4}
    z=x/\epsilon^2 \,, \quad \tau=t/\epsilon \, ,
\end{equation}
in which the spin relaxation time $1/\alpha$ is selected as the unit of time instead of $1/\omega_0$ as in table \ref{t1}. This choice is the natural one to monitor spin relaxation at high temperature. In Fig. \ref{fig2}, we observe that the linear approximation breaks down for temperatures closer or equal to the critical value $\theta=1$ but the nonlinear equation is still a good approximation of the simulation values. In particular, this is the situation in the overdamped region $\theta_d^-<\theta<\theta_d^+$, where $\theta_d^\pm$ are given by (\ref{4.17}) and (\ref{4.22}), respectively. For the value $\epsilon=0.25$, $\theta_d^-=0.998$ and $\theta_d^+=1.004$. It is a very narrow region, being its width of order $\epsilon^2$. Therefore, the evolution of the oscillator is almost indistinguishable from the critical temperature behaviour, shown in figure \ref{fig2}(b).

\begin{figure}[htbp]
\begin{center}
\includegraphics[width=6cm]{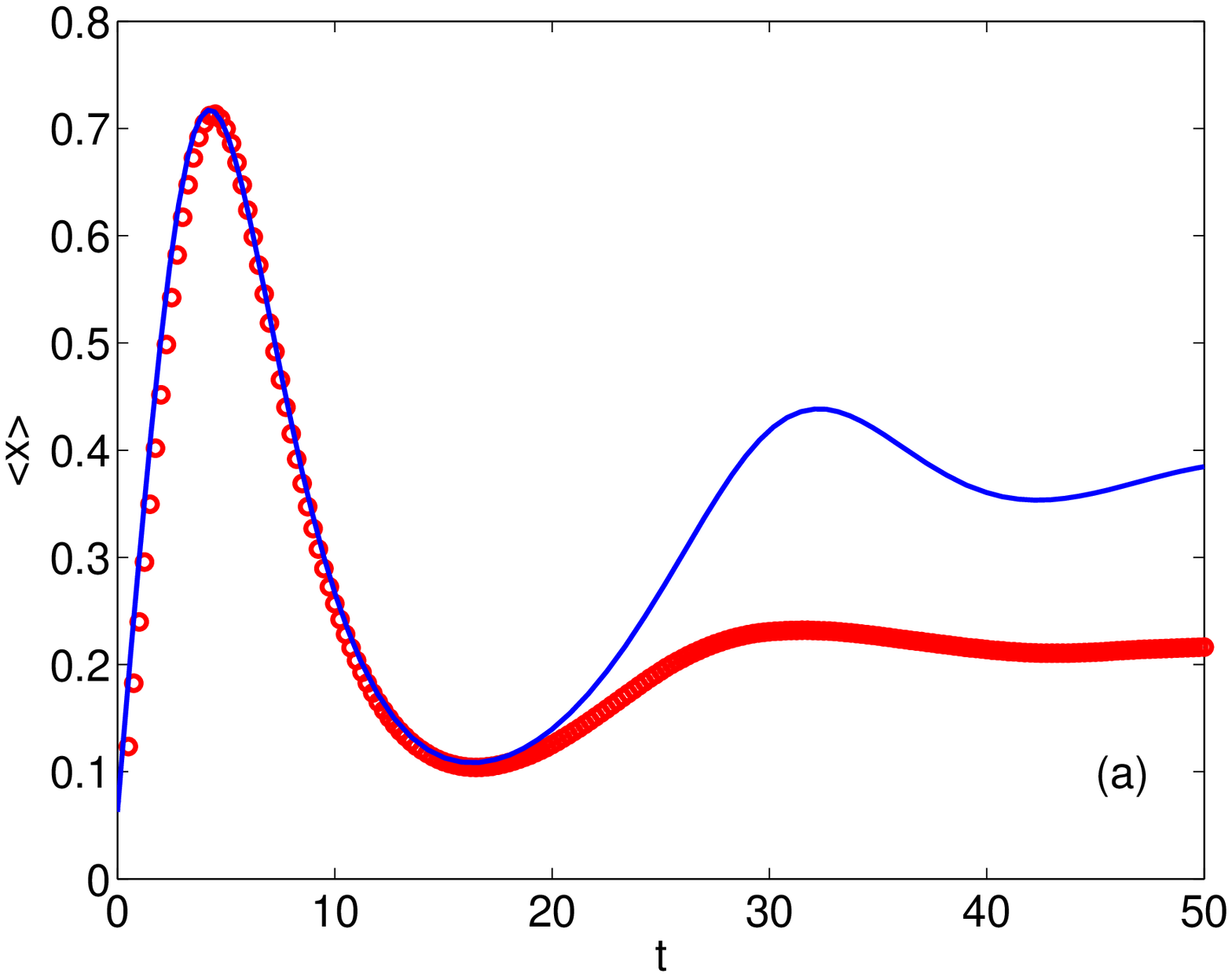}
\includegraphics[width=6cm]{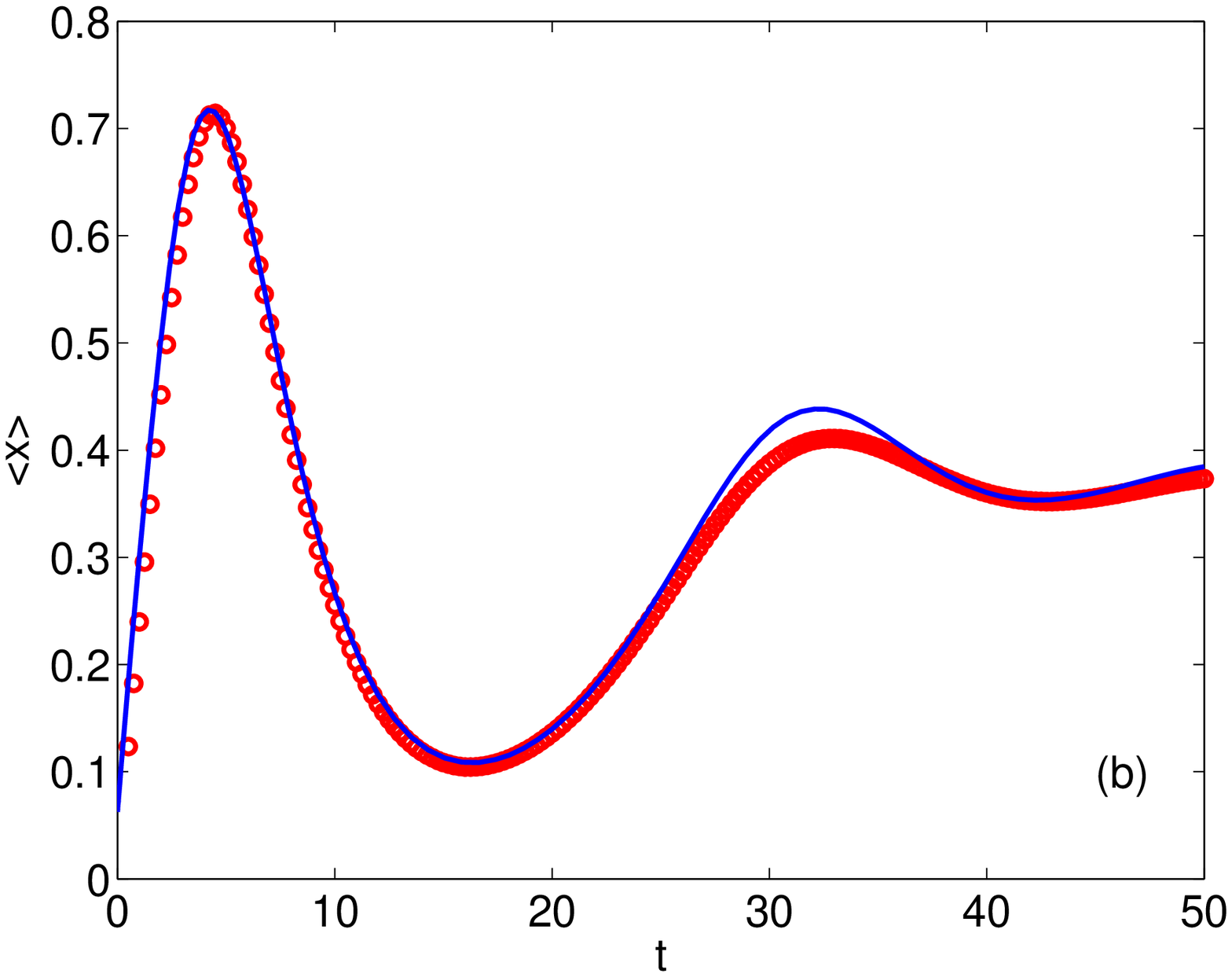}
\includegraphics[width=6cm]{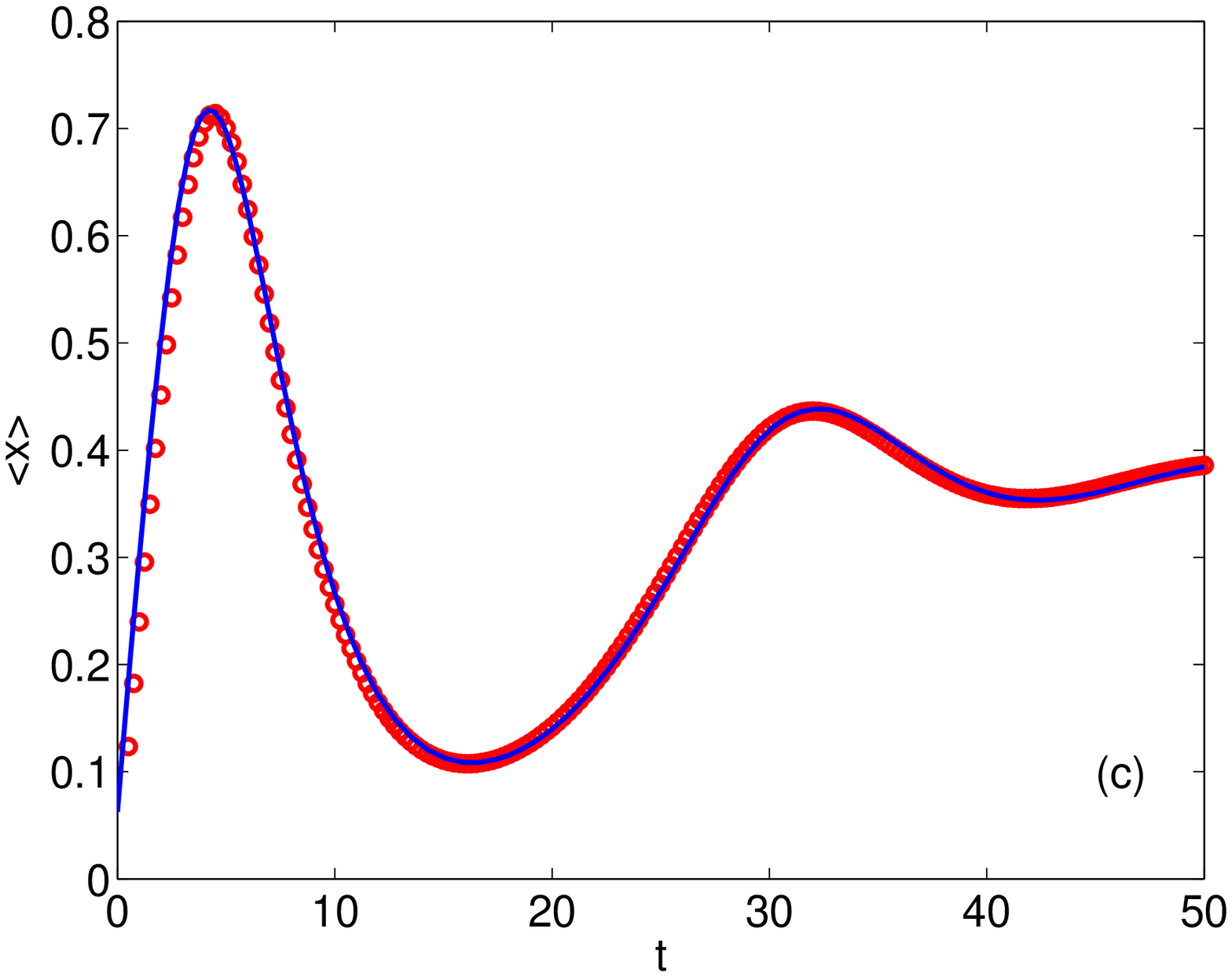}
\caption{Same as Figure \ref{fig4} for $\theta=0.95$ (below the critical temperature) but calculated with different number of spins and trajectories: (a) $N=10^4$, $N_T=10^2$;  (b) $N=10^5$, $N_T=10^2$; (c) $N=10^6$, $N_T=40$ (in fact $N_T=5$ suffices).
Initial values $({\epsilon^2},{\epsilon})$. }
\label{fig5}
\end{center}
\end{figure}

\begin{figure}[htbp]
\begin{center}
\includegraphics[width=9cm]{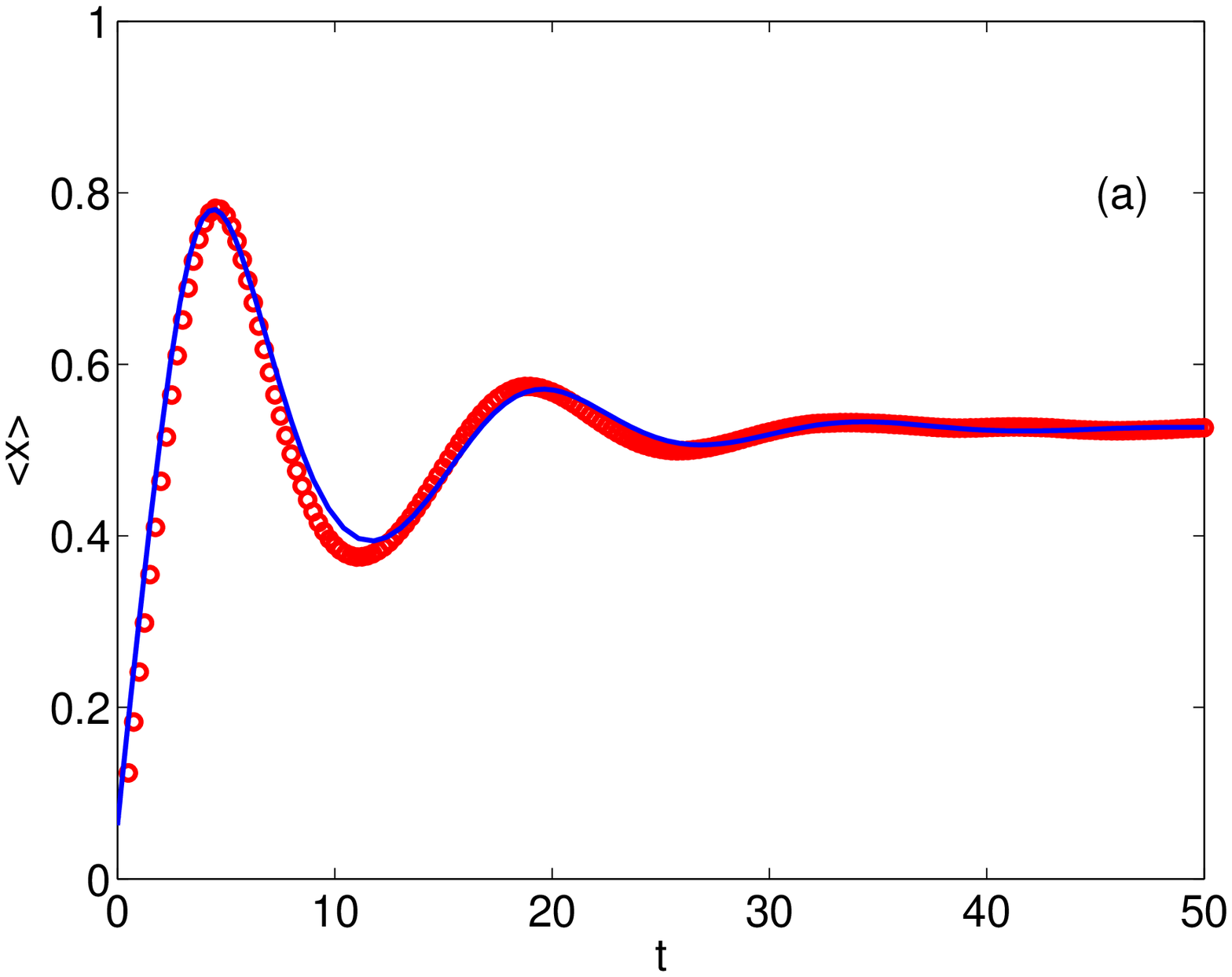}
\includegraphics[width=9cm]{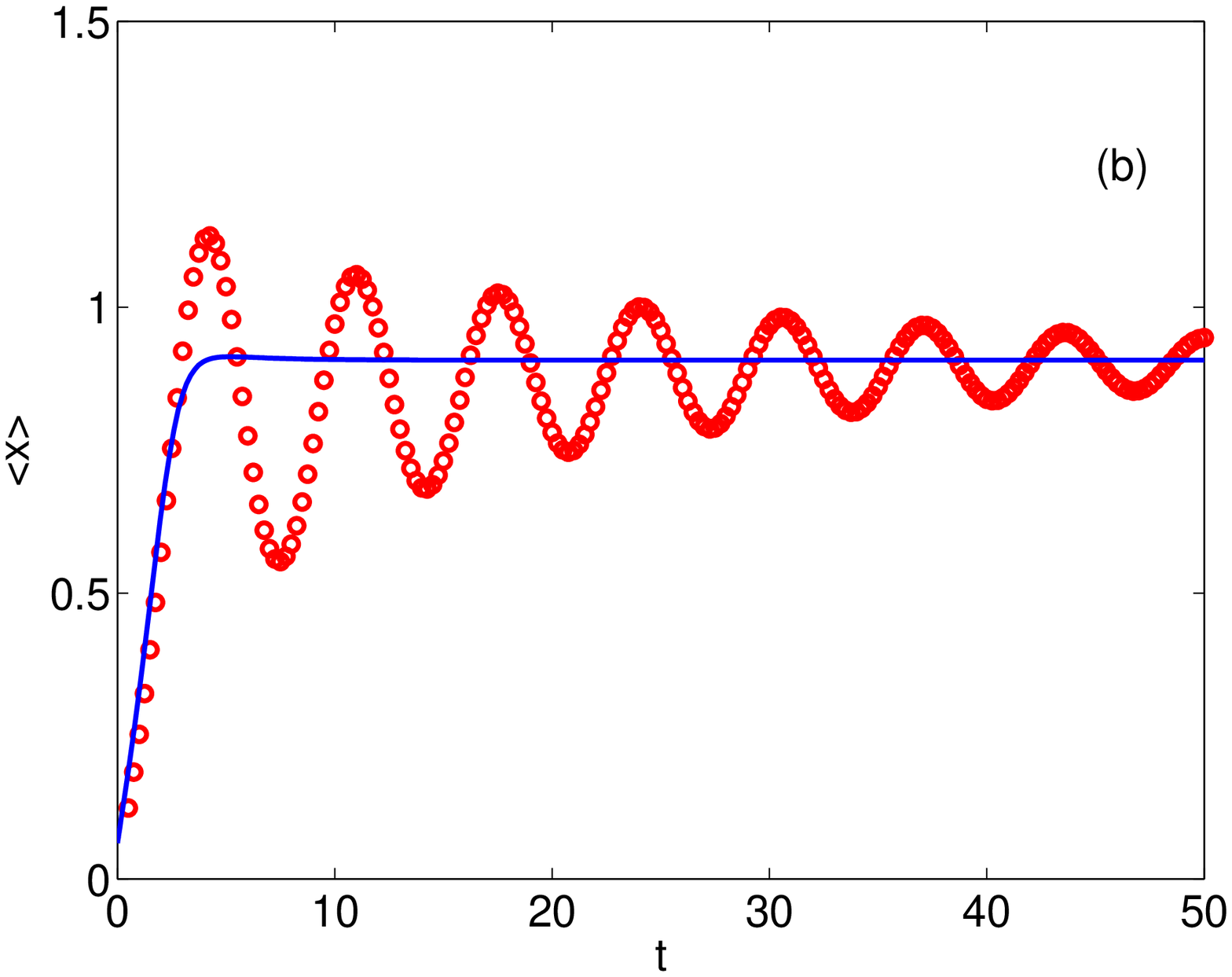}
\caption{Averaged trajectories $\langle x(t)\rangle=\widetilde{x}(t)$ (circles) versus the
nonlinear (solid blue line) prediction for
initial data $\widetilde{x}(0)=\epsilon^2$, $\dot{\widetilde{x}}(0)=\epsilon$ and (a) $\theta=0.9$ and (b) $\theta=0.6$. Other parameter values are as in Figure \ref{fig1}.}
\label{fig6}
\end{center}
\end{figure}

Figures \ref{fig5} depicts the evolution of the oscillator position toward one of the two nonzero equilibrium values for $\theta=0.95$, below the critical temperature. To attain good agreement between the prediction of the nonlinear equation and the averages over trajectories, the number of particles in our simulations has to increase while a small number of trajectories (as low as 5) suffices: compare Fig. \ref{fig5}(a) for $N=10^4$ with \ref{fig5}(b) for $N=10^5$ and with \ref{fig5}(c) for $N=10^6$. For $\theta=0.95$, Figure \ref{fig5}(c) shows that the nonlinear approximation gives again a good description of the oscillator dynamics, accounting for its time evolution to $\widetilde{x}_{\text{eq}}\simeq 0.38$, as predicted by (\ref{3.2}). If the temperature is further lowered, the number of spins necessary for the prediction of the nonlinear equation to approximate the simulation values again decreases to $N=10^4$ with $N_T=100$ trajectories. This can be seen in figure \ref{fig6}(a) for $\theta=0.9$. For $\theta=0.6$, figure \ref{fig6}(b) shows that the nonlinear equation predicts a monotonic approach to the equilibrium value $\widetilde{x}_{\text{eq}}\simeq 0.91$. On the other hand, the simulation gives underdamped oscillations towards $\widetilde{x}_{\text{eq}}$, with a period approximately given by the oscillator natural period. Similar curves are found for lower temperatures. The nonlinear equation has the correct equilibrium oscillator positions as stable stationary solutions (so it gives the attractors correctly), but is not expected to be accurate for temperatures below that given by  (\ref{4.13b}). Estimating $\widetilde{x}$ by its steady value (\ref{3.3}), the lowest temperature for which the nonlinear equation (\ref{4.11}) is expected to hold is given by $\theta=0.877$. Let us recall that the reason for this is that the spin relaxation time diverges as $T\rightarrow 0$ \cite{Re80,ByP93b,ByP96}, and the separation of time scales leading to (\ref{4.11}) is no longer valid.

\section{Conclusions}
\label{s6}

We have studied a harmonic oscillator subject to a force due to a chain of spins whose coupling constant is proportional to the oscillator position. The spins are in contact with a thermal bath at constant temperature and evolve following Glauber's dynamics. We have shown that the oscillator potential energy is modified by the spins and that it experiences a nonlinear friction. The quasi-stationary approximation (\ref{4.10}) is basically a linear theory around equilibrium, which is valid if $\epsilon\ll 1$. Physically, this means that the natural oscillator period $2\pi/\omega_0$ is much larger than the characteristic relaxation time of the  spins'  energy. Then the spins relax to equilibrium over a time scale in which the position of the oscillator can be considered roughly constant. The equilibrium contribution of $\widetilde{C_1}$ accounts for the renormalization of the potential, while the term corresponding to the (small) deviation of the Ising system from this ``equilibrium'' gives rise to the friction term.

The oscillator rest points are the stationary solutions of the corresponding reduced dynamical equation. These solutions undergo a supercritical pitchfork bifurcation as the bath temperature crosses a critical value. For temperatures above critical, the stable equilibrium position of the oscillator is zero, the same as that of the uncoupled oscillator. Below the critical temperature, there are two stable symmetric equilibrium positions. This pitchfork bifurcation corresponds to a second order phase transition for the equilibrium probability of the oscillator-spin system. The oscillator equilibrium position is the corresponding order parameter and it plays the same role as the magnetization in an effective long range 1d Ising system.

Even when our dynamical equation (\ref{4.11}) does not give an accurate description of the oscillator time evolution for very low temperatures, the equilibrium points are always correctly predicted by the solutions of (\ref{3.2}). This is not surprising: (\ref{3.2}) is exact in the thermodynamic limit, independently of the value of $\epsilon$, while (\ref{4.11}) holds only if the characteristic relaxation time of the spins is much smaller than the oscillator natural period. This condition is not fulfilled for $T\rightarrow 0$, because the spin relaxation time diverges in that limit \cite{Re80,ByP93b,ByP96}.

\ack
The authors thank J.\ Javier Brey for carefully reading the manuscript and providing useful comments.
This research has been supported by the Spanish Ministerio de Ciencia e Innovaci\'on (MICINN) through Grants No. FIS2008-01339 (AP, partially financed
by FEDER funds), FIS2008-04921-C02-01 (LLB), and FIS2008-04921-C02-02 (AC). The authors would like also to thank the Spanish National Network Physics of Out-of-Equilibrium Systems financed through the MICINN grant FIS2008-04403-E.

\setcounter{equation}{0}
\renewcommand{\theequation}{A.\arabic{equation}}

\appendix
\section{Normal solution of the system of equations (\ref{4.8}) - (\ref{4.9})}
\label{apa}

The solution of (\ref{4.8}) satisfying $\widetilde{C^{(0)}_0}=1$ is
\begin{eqnarray}
&& \widetilde{C^{(0)}_{n}}= \eta^n, \quad \eta=\tanh\left(\frac{\widetilde{x}}{\theta}\right) \, \label{a.1}\\
&& \gamma=\frac{2\eta}{1+\eta^2} \,, \label{a.2}
\end{eqnarray}
in which $\widetilde{C^{(0)}_n}$ is bounded as $n\to\infty$. Inserting these expressions in (\ref{4.9}), we get
\begin{equation}
\frac{2\eta}{1+\eta^2} \left(\widetilde{C^{(1)}_{n-1} }+
\widetilde{ C^{(1)}_{n+1} }\right)-2 \widetilde{ C^{(1)}_n }= n\eta^{n-1}\frac{\rmd \eta}{\rmd t},
\label{a.3}
\end{equation}
with the boundary condition $\widetilde{C^{(1)}_0}=0$. This equation can be solved
by using standard methods for difference equations \cite{Be99}, with the result
\begin{equation}\label{a.4}
  \widetilde{C^{(1)}_n}= a_n \eta^n \, ,
\end{equation}
where
\begin{equation}\label{a.5}
    a_n=\sum_{i=0}^{n-1} b_i \, , \qquad  b_n=-\frac{1}{2\eta}\frac{1+\eta^2}{1-\eta^2}
    \frac{\rmd\eta}{\rmd t} \left(n+\frac{1}{1-\eta^2}\right)  \, .
\end{equation}
Then $\widetilde{C_1}\sim (1+ a_1 \epsilon)\eta$ yields (\ref{4.10}).

\end{document}